\documentclass[acmlarge]{acmart}%IMWUT

\AtBeginDocument{%
 \providecommand\BibTeX{{%
 \normalfont B\kern-0.5em{\scshape i\kern-0.25em b}\kern-0.8em\TeX}}}

\setcopyright{acmlicensed}
\acmJournal{IMWUT}
\acmYear{2021} \acmVolume{1} \acmNumber{1} \acmArticle{1} \acmMonth{1} \acmPrice{15.00}\acmDOI{10.1145/3494976}

\usepackage{multirow}
\usepackage{colortbl}

\newlength\savedwidth
\newcommand{\wcline}[1]{\noalign{\global\savedwidth\arrayrulewidth\global\arrayrulewidth 1.0pt} \cline{#1}
\noalign{\global\arrayrulewidth\savedwidth}}%thickness of lines in Table

\newcommand{\HL}[1]{\textcolor{black}{#1}}

\begin{document}

\title[Computing Touch-Point Ambiguity on Mobile Touchscreens for Modeling Target Selection Times]{Computing Touch-Point Ambiguity on Mobile Touchscreens for Modeling Target Selection Times}

\author{Shota Yamanaka}
\affiliation{%
 \institution{Yahoo Japan Corporation}
 \city{Chiyoda-ku}
 \state{Tokyo}
 \country{Japan}
}
\email{syamanak@yahoo-corp.jp}

\author{Hiroki Usuba}
\affiliation{%
 \institution{Meiji University}
 \city{Nakano-ku}
 \city{Tokyo}
 \country{Japan}
}

\renewcommand{\shortauthors}{Yamanaka and Usuba}

\begin{abstract}
Finger-Fitts law (FFitts law) is a model to predict touch-pointing times, modified from Fitts' law.
It considers the absolute touch-point precision, or a \textit{finger tremor factor} $\sigma_\mathrm{a}$, to decrease the admissible target area and thus increase the task difficulty.
Among choices such as running an independent task or performing parameter optimization, there is no consensus on the best methodology to measure $\sigma_\mathrm{a}$.
\HL{This inconsistency could be detrimental to HCI studies such as pointing technique evaluations and user group comparisons.}
By integrating the results of our 1D and 2D touch-pointing experiments and reanalyses of previous studies' data, we examined the advantages and disadvantages of each approach to compute $\sigma_\mathrm{a}$.
We found that the parameter optimization method is a suboptimal choice for predicting the performance.
\end{abstract}


\maketitle

\section{Introduction}
\subsection{Background}
For human motor performance modeling, researchers have sought to develop new models and modify existing models to improve their prediction accuracy (i.e., model fitness).
The model that we focus on here, the Finger-Fitts law (a.k.a., FFitts law) proposed by Bi et al. \cite{Bi13a}, is a modified version of Fitts' law \cite{Fitts54} for predicting operational times in target pointing on touchscreens.
FFitts law is based on the effective width method \cite{Crossman56}, which adjusts the target size (or width $W$) from the \textit{nominal} value drawn on the screen to an \textit{effective} width that takes the actual touch-point distributions into account.
Bi et al. modified this effective width method to deal with finger touch ambiguity and empirically showed that FFitts law is superior to Fitts' law in terms of model fitness \cite{Bi13a}.
As touchscreen devices have become common in our daily life, deriving a model with a high prediction accuracy will contribute directly to HCI, e.g., when designers create user interfaces for webpages and apps.

As another type of contribution in performance modeling, standardizing a model's methodology is important for future researchers in terms of the replicability \cite{MacKenzie92,Soukoreff04,Wobbrock11dim}. 
Unfortunately, while several research groups have examined FFitts law \cite{Bi13a,Ko20FF2D,Noort15run,Yamanaka18iss,Yamanaka18mobilehci,Woodward20avi}, there is no consensus on a standard methodology, which is an obstacle to future research on finger-touch pointing.
The methodology inconsistencies include the computation method for touch ambiguity, the instruction for the finger calibration task, and the target size for the calibration task.
There are two other issues of inconsistency: in contrast to Bi et al.'s finding \cite{Bi13a}, the model fitness of FFitts law was also found to be inferior to that of Fitts' law with \HL{the} nominal width, while FFitts law was superior to Fitts' law with \HL{the} effective width \cite{Woodward20avi}; and FFitts law sometimes cannot be used because of a mathematical error that results when the value inside a square root (sqrt) is negative \cite{Yamanaka18mobilehci,Yamanaka18iss}.

\HL{Leaving these inconsistent methodologies and issues unresolved could be detrimental to HCI studies such as evaluation of novel pointing techniques and comparison of different user groups.}
This point was previously mentioned in regard to Fitts' law \cite{Soukoreff04}, and rethinking the finger-touch model (FFitts law) is a timely notion given the recent trend of widespread smartphone and tablet use.
In this paper, taking a step toward a standard for measuring touch-pointing performance, we explain the concept of FFitts law and survey the inconsistencies of its methodologies in the literature.
Then, we empirically examine how the inconsistent methodologies change the results of FFitts law for both 1D and 2D target-pointing tasks, through eight sub-tasks in total.
Our contributions are twofold.
\begin{itemize}
\setlength{\leftskip}{-5mm}%indent
\item We survey related work on FFitts law to explore inconsistencies in its methodologies (Section~\ref{sec:predis}), and we reanalyze previous FFitts law studies with modern methods (Section~\ref{sec:reanalysis}). These sections emphasize that previous researchers have run different procedures for a single model, and the relative advantages and disadvantages of each approach. For example, parameter optimization can always be applied (i.e., to avoid a negative value inside the \HL{square} root), but it induces the risk of overfitting the data.
\item We conduct eight sub-tasks in total, including two main Fitts' law tasks with 1D and 2D targets. The results show that the baseline Fitts' law model has the highest fitness, with adjusted $R^2 = 0.98$, and that the parameter optimization method with the nominal target width has comparable prediction accuracy according to the information criteria and cross-validation. Our results and reanalyses suggest that using an extra free parameter for the finger tremor does not cause a critical overfitting problem and can yield better fits in some cases.
\end{itemize}

\subsection{Motivations and Implications Related to Ubiquitous Touch Device Usage}

Smartphones and tablets are often used in a static situation such as a user standing or sitting on a chair, as in the experiments described in this paper.
In addition, because of the mobility of smart devices, users often operate them while walking or running.
In such non-static conditions, as the walking speed increases, the touch-pointing performance deteriorates in terms of operational times and error rates \cite{Bergstrom11walking,Lin07walking,Schildbach10walking}.
Under running conditions, this performance reduction is even more clearly observed \cite{Noort15run}.
Moreover, touchscreens besides smartphones and tablets are becoming ubiquitous, and similar issues have been reported for those cases.
For example, under vibration conditions, touch-pointing performance is degraded for car navigation systems \cite{Ahmad15carnavi,Tao21vib} and cockpit touch displays \cite{Cockburn17cockpit,Coutts19cockpit,Zon20cockpit}.

Because Fitts' law is a basis for designing better UIs, our work will enable researchers and practitioners to do so for touch-based systems.
For example, without a standardized methodology to apply FFitts law, designers have to conduct (potentially costly) user studies to determine a suitable target size for rapidly tapping a button within (e.g.) 800~ms when a car is driven at 20~km/h, 60~km/h, 100~km/h, and so on.
In addition, a standardized methodology will increase the reliability of data predicted by a model, which will enable the model to help optimize UIs \cite{Bailly13,Eggers03} and generate user-friendly UIs automatically \cite{Gajos04,Nichols04}.

Although user experiments were conducted in the previous studies mentioned above, the choices of experimental tasks in those studies do not generalize to other untested conditions.
For example, in a touch-pointing experiment with users walking \cite{Schildbach10walking}, three square target sizes of 6.74, 8.18, and 9.50 mm for each edge were tested.
Then, by comparing the standing and walking situations, the authors concluded that operational times while walking were longer than those while standing, in particular for smaller targets.
The benefits of user performance models mean that we can estimate the potential decrease in operational times for untested conditions, such as a 5-mm target with users walking, without conducting additional user studies.
Because finger-touch ambiguity is a critical factor for precise touch operations in non-static conditions, the need for a robustly applicable FFitts law methodology will be even more important for designing better UIs related to the future ubiquity of touch devices.
Still, regarding the external validity of our conclusions, such as whether the parameter optimization method is effective for a vibration situation in a car or during walking, we will need further empirical evidence, which we will obtain in our future work.

\section{Related Work}
\subsection{Fitts' Law and the Effective Width Method}
According to Fitts' law, the movement time $\mathit{MT}$ for pointing is linearly related to the index of difficulty, $\mathit{ID}$ \cite{Fitts54}:
\begin{equation}
\label{eqn:IDfitts}
\mathit{MT}=a+b\cdot{}\mathit{ID},
\end{equation}
where $a$ and $b$ are constants.
In the HCI field, the Shannon formulation is widely used for the $\mathit{ID}$ value \cite{MacKenzie92}:
\begin{equation}
\label{eqn:ID_n}
\mathit{ID}_{n}=\log_{2}\left(A/W+1\right),
\end{equation}
where $A$ is the distance to the target and $W$ is its width.
Here, $A$ and $W$ are {\sl nominal} values shown on the display.

In typical pointing experiments, participants are instructed to ``point to a target as rapidly and accurately as possible'' \cite{Soukoreff04}.
However, it is common that some participants tend to show short $\mathit{MT}$ values and high error rates, while others show long $\mathit{MT}$ values and low error rates \cite{Zhai04speed}.
To normalize such biases in comparing those participants' performance, or to compare the performance with different input devices (e.g., finger vs. stylus), using Crossman's post-hoc correction for calculating the effective target width $W_e$ \cite{Crossman56} is recommended \cite{MacKenzie92,Soukoreff04,Wobbrock11dim}:
\begin{equation}
\label{eqn:W_e}
W_e=\sqrt{2 \pi e}\sigma_\mathrm{obs} = 4.133\sigma_\mathrm{obs},
\end{equation}
where $\sigma_\mathrm{obs}$ is the standard deviation $\mathit{SD}$ of the \textbf{\underline{obs}}erved endpoints.
This adjustment is based on an assumption that the spread of hits follows a normal distribution.
By using this method, the $W_e$ is adjusted so that $\sim$96$\%$ of hits fall inside the target.
The effective $\mathit{ID}$ using the $W_e$ is defined as $\mathit{ID}_{e}=\log_{2}\left(A/W_e+1\right)$.
\HL{While Gori et al. \cite{Gori18tochi} questioned the theoretical justification of the effective width method produced in \cite{Soukoreff04}, they also provided some support for $\mathit{ID}_e$; see Section 7.1 in \cite{Gori20BC}.}
FFitts law is also based on this effective width method.

\subsection{Overview of Finger-Fitts Law}
Bi, Li, and Zhai hypothesized that the observed spread of hits ($\sigma_\mathrm{obs}$) includes \HL{both relative and absolute components:} the former component follows the speed-accuracy tradeoff rule, while the latter one solely depends on the finger touch precision \cite{Bi13a}.
The tapped point is considered a random variable $X$ following a normal distribution ($X \sim N(\mu, \sigma_\mathrm{obs} ^2)$).
Then, $X$ is the sum of two independent random variables for the \textbf{\underline{r}}elative and \textbf{\underline{a}}bsolute components, both of which follow normal distributions: $X_\mathrm{r} \sim N(\mu_\mathrm{r}, \sigma_\mathrm{r}^2)$ and $X_\mathrm{a} \sim N(\mu_\mathrm{a}, \sigma_\mathrm{a}^2)$, respectively.
\HL{Bi et al. called this ``the dual Gaussian distribution hypothesis.''}
Although the relative spread of hits, $\sigma_\mathrm{r}$, decreases as the movement speed and target width decrease, the absolute finger precision $\sigma_\mathrm{a}$ cannot be controlled via a user's speed-accuracy priority.
The means of both components ($\mu_\mathrm{r}$ and $\mu_\mathrm{a}$) are assumed to be close to the target center: $\mu_\mathrm{r} = \mu_\mathrm{a} = 0$.

Here, $\sigma_\mathrm{r}$ is what the effective width method models.
Thus, from Equation~\ref{eqn:W_e}, Bi et al. \cite{Bi13a} derived
\begin{equation}
\label{eqn:We_r}
W_e=\sqrt{2 \pi e}\sigma_\mathrm{r}.
\end{equation}
Because Bi et al. assumed that $X$ is the sum of the independent random variables $X_\mathrm{r}$ and $X_\mathrm{a}$, $\sigma_\mathrm{obs}^2$ is written as
\begin{equation}
\label{eqn:sigma_sum}
\sigma_\mathrm{obs}^2 = \sigma_\mathrm{r}^2 + \sigma_\mathrm{a}^2.
\end{equation}
From Equations~\ref{eqn:We_r} and \ref{eqn:sigma_sum}, the effective width for FFitts law, $W_f$, is derived as
\begin{equation}
\label{eqn:Wf_Bi}
W_f=\sqrt{2 \pi e (\sigma_\mathrm{obs}^2 - \sigma_\mathrm{a}^2)}.
\end{equation}

\subsection{Measurement of the Touch Ambiguity Factor \texorpdfstring{$\sigma_\mathrm{a}$}{sigma a}}
\subsubsection{Finger Calibration Task with ``Rapid and Accurate'' Instruction}
Bi et al. obtained $\sigma_\mathrm{a}$ via 1D and 2D finger calibration tasks conducted independently from the Fitts' law task \cite{Bi13a}.
In the 1D task, participants repeatedly tapped as closely to a 2.4-mm-high horizontal bar target as possible, and the $\mathit{SD}$ of the signed biases from the target was computed as $\sigma_\mathrm{a}$.
For the 2D condition, a 2.4-mm-diameter circle was used as the target, and the bivariate $\mathit{SD}$ was taken as $\sigma_\mathrm{a}$.
In both tasks, the participants were instructed to tap the target as rapidly and accurately as possible.
Because this task does not require a movement to a target from a specific position, Bi et al. stated that the speed-accuracy tradeoff rule has a negligible effect on $\sigma_\mathrm{a}$.

Woodward et al. conducted FFitts law tasks with children and circular targets \cite{Woodward20avi}.
Overall, they followed the procedure of Bi et al.
For the calibration task, they used a target with $W=4.8$ mm, which was also the smallest size for the main Fitts' law task.
Their paper does not explicitly state whether the participants were instructed to balance speed and accuracy or to concentrate on accuracy.

\subsubsection{Finger Calibration Task with ``Concentrate on Accuracy'' Instruction}
Luo and Vogel tested the applicability of FFitts law to touch-based goal-crossing tasks \cite{Luo14}.
They drew a 2-pixel line for the finger calibration task and instructed the participants ``not to rush and focus on accuracy,'' because ``measuring $\sigma_\mathrm{a}$ is not about speed.''
Hence, in contrast to Bi et al.'s instruction, Luo and Vogel removed the instruction of ``operating as rapidly as possible.''
They reported somewhat negative results: the data fit for the discrete crossing condition decreased from $R^2 = 0.7526$ (conventional Fitts' law) to $0.5853$ (FFitts law).
After removing the data point with the highest $\mathit{ID_n}$, the FFitts law fitness improved to $R^2=0.843$, but this was likely due to an arbitrary choice of data-point removal to increase $R^2$.

Yamanaka tested Fitts' and FFitts laws for touch-pointing tasks with unwanted target items (called \textit{distractors}) \cite{Yamanaka18mobilehci,Yamanaka18iss}.
In the finger calibration tasks, 1-pixel targets were used (a bar for 1D and a crosshair for 2D).
As in Luo and Vogel's study, Yamanaka instructed the participants to ``tap as close to the target as possible'' and emphasized that the ``participants were instructed to concentrate on spatial precision and not on time.''
He reported that FFitts law could not be used, because in some task conditions, the $\sigma_\mathrm{obs}$ values were smaller than $\sigma_\mathrm{a}$, resulting in a negative value inside the square root in FFitts law (Equation~\ref{eqn:Wf_Bi}).
This mathematical error occurred even in no-distractor conditions (i.e., {a typical Fitts task}).

\subsubsection{Intercept of Regression between the Squares of $\sigma_\mathrm{obs}$ and $W$}
\begin{figure}[t]
\centering
\includegraphics[width=0.90\columnwidth]{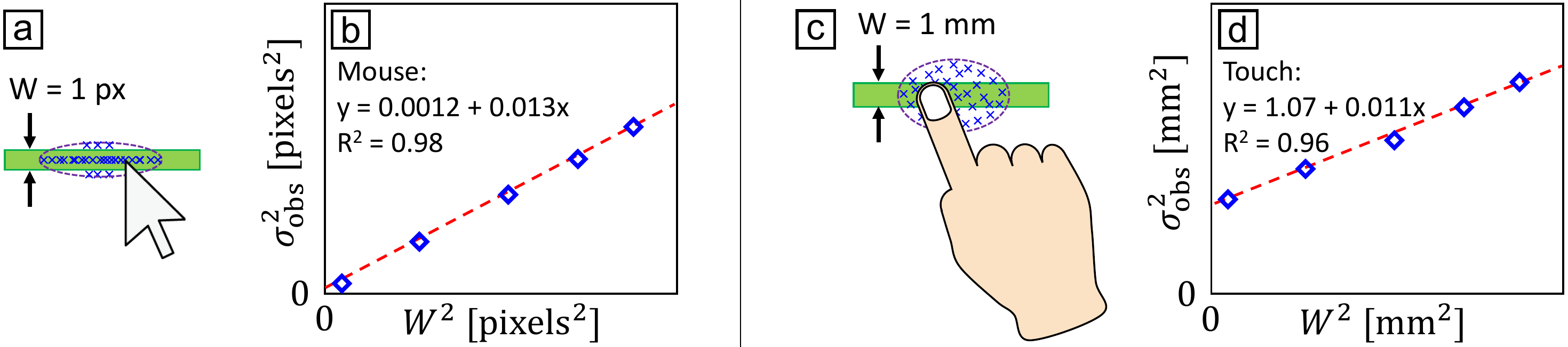}
\caption{Overview of the dual Gaussian distribution hypothesis with hypothetical endpoint data (the `X' marks). (a) In a mouse-pointing task, if a user spends a sufficient time to point to an extremely small target, the observed endpoint variability $\sigma_\mathrm{obs}$ on the y-axis is close to zero; thus, (b) the intercept in the regression of the squares of $\sigma_\mathrm{obs}$ and $W$ is also close to zero. (c) In contrast, for touch pointing, even if a user spends a long time, there is a remarkable variability when tapping a small target \cite{Holz10,Holz11}; thus, (d) the regression has a clear nonzero intercept.}
\label{fig:overview-sigma-w}
\end{figure}

In Bi and Zhai's 2D touch-pointing task, at the beginning of each trial, a circular target appeared on the screen, and the participants tapped it as rapidly and accurately as possible \cite{Bi13b}.
Bi and Zhai assumed that the endpoints when using a fine probe like a mouse cursor are proportionally related to $W$ (i.e., $\sigma_\mathrm{r} = \mathrm{constant} \times W$), thus giving
\begin{equation}
\label{eqn:sigma2_W2}
\sigma_\mathrm{r}^2 = \alpha W^2,
\end{equation}
where $\alpha$ is a constant.
By substituting this $\sigma_\mathrm{r}^2$ from Equation~\ref{eqn:sigma2_W2} into Equation~\ref{eqn:sigma_sum}, we obtain
\begin{equation}
\label{eqn:sigma2_a2_W2}
\sigma_\mathrm{obs}^2 = \alpha W^2 + \sigma_\mathrm{a}^2.
\end{equation}
Figure~\ref{fig:overview-sigma-w} shows this relationship.
They used five circular target diameters ($W$ = 2, 4, 6, 8, and 10 mm), and their regression expression for $W^2$ versus the corresponding $\sigma_\mathrm{obs}^2$ values on the (e.g.) y-axes gave $\sigma_\mathrm{obs}^2 = 0.0108 W^2 + 1.3292$.
From this, $\sigma_\mathrm{a}$ was computed as $\sqrt{1.3292} = 1.153$ mm.

\subsubsection{Parameter Optimization}
Ko et al. proposed to obtain the finger tremor factor by parameter optimization \cite{Ko20FF2D}.
They indicated that Equation~\ref{eqn:Wf_Bi} can be rewritten as follows, according to their Equation 4 on p. 859 \cite{Ko20FF2D}:
\begin{equation}
\label{eqn:We_c_ko_1st}
W_f = \sqrt{2 \pi e (\sigma_\mathrm{obs}^2 - \sigma_\mathrm{a}^2)} = \sqrt{2 \pi e \sigma_\mathrm{obs}^2 - 2 \pi e \sigma_\mathrm{a}^2} = \sqrt{ W_e^2 - 2 \pi e \sigma_\mathrm{a}^2}.
\end{equation}
Then, they made two simplifications.
First, they used the nominal $W$ instead of $W_e$, which ``assumes that participants respect the spatial constraint set by the task parameters'' (p. 860).
Second, they replaced $2 \pi e \sigma_\mathrm{a}^2$ in Equation~\ref{eqn:We_c_ko_1st} with an empirically determined free parameter $c^2$, which assumes that ``the absolute error caused by finger [$\sigma_\mathrm{a}$] varies in different task contexts'' (p. 861).
Therefore, the model on $MT$ is written as:
\begin{equation}
\label{eqn:We_c_ko}
\mathit{MT} = a+b\cdot{}\log_2 \left( \frac{A}{ \sqrt{ W^2 - c^2}} +1 \right).
\end{equation}
We use this as a candidate model.
Ko et al.'s original goal was to model rectangular-target pointing on touchscreens \HL{for which} a target's width and height are defined as $W$ and $H$, respectively.
One of their models uses the smaller of $W$ and $H$ as the target size, which was proposed in previous studies \cite{Hoffmann94height,MacKenzie92twoD}, as follows:
\begin{equation}
\label{eqn:smallerOf}
\mathit{MT} = a+b\cdot{}\log_2 \left( \frac{A}{ \sqrt{ \mathrm{min}(W,H)^2 - c^2}} +1 \right).
\end{equation}
If we use circular targets whose size is solely defined by $W$, this model \HL{is equivalent to} Equation~\ref{eqn:We_c_ko}.
Therefore, we should note that Equation~\ref{eqn:We_c_ko} is a special case of Ko et al.'s model.
In their rectangular-target pointing task, they empirically confirmed that using the nominal $W$ instead of $W_e$ gave a higher model fitness, which is consistent with previous studies on the effective width method (e.g., \cite{Wright13,Zhai04speed}).

In fact, Equation~\ref{eqn:We_c_ko} using $W_e$ was proposed by Welford in 1968 (p. 156, l.30 in \cite{Welford68}) with the ``+0.5'' version of Fitts' law instead of ``+1''.
The +0.5 version has also been used in the HCI field \cite{Soukoreff04}.
\HL{Welford's aim was the same: $c$ represents a hand tremor in a stylus-tapping task.}
Also, he empirically confirmed that the following ``no root, no power'' formulation showed a better fit than using $\sqrt{ W^2 - c^2}$ (note that he examined $W_e$ instead of $W$):
\begin{equation}
\label{eqn:We_c_welford}
\mathit{MT} = a+b\cdot{}\log_2 \left( \frac{A}{ W - c} +1 \right).
\end{equation}
This model's superiority with respect to the baseline (Equation~\ref{eqn:ID_n}) for small targets was confirmed by Chapuis and Dragicevic's mouse pointing tasks \cite{Chapuis11}.
They also found that this model using $W_e$ was superior to the original effective width method ($\mathit{ID_e} = \log_2 (A/W_e+1)$).
They estimated the $c$ value from the empirical data (i.e., parameter optimization), and they also reported that the $c$ varied among different tasks.

Because our purpose in this study was to examine how the conclusions would change depending on the different methodologies and models, we sought to compare all of the candidate model formulations found in the literature.
Hence, when we began this work (November 2020), we surveyed all research papers that cited Bi et al.'s FFitts law paper \cite{Bi13a} in the ACM Digital Library and Google Scholar, and we examined the potential FFitts law formulations.

\section{Discussion on Inconsistencies and Problems of FFitts Law}
\label{sec:predis}
\subsection{Target Size in Calibration Task}
\label{sec:calSize}
There are two kinds of approaches: using the smallest $W$ used in a Fitts' law task (2.4-mm \cite{Bi13a} or 4.8-mm target \cite{Woodward20avi}) or the minimum visible target (1 pixel \cite{Yamanaka18mobilehci,Yamanaka18iss} or 2 pixels \cite{Luo14}).
In pointing tasks with a fine probe, $\sigma_\mathrm{obs}$ is assumed to be proportional to $W$ when users can spend sufficient time.
In this case, users can accurately point to a small target even if the width is quite narrow: e.g., $W=1$ pixel.
In contrast, in touch-pointing tasks, there is an unavoidable lower bound on the finger precision $\sigma_\mathrm{a}$.
Hence, even if users can spend a long time, there is a slight distance from the intended target position to the actual tapped position sensed by the system \cite{Bi13a,Holz11}.
The aim of a finger calibration task is to measure this lower bound of precision as the variance of the tapped position in the Fitts' law paradigm.
For this purpose, pointing to a 1-pixel target with the instruction to operate as rapidly and accurately as possible is a straightforward method.

There is an issue related to using the smallest $W$ in the main Fitts' law task.
The issue is that we may observe a mathematical error in the \HL{square} root in Equation~\ref{eqn:Wf_Bi} ($W_f=\sqrt{2 \pi e (\sigma_\mathrm{obs}^2 - \sigma_\mathrm{a}^2)}$).
For example, Woodward et al. used a target with $W=4.8$ mm for the calibration, and the observed $\mathit{SD}$ ($=\sigma_\mathrm{a}$) was 1.590148 mm \cite{Woodward20avi}.
The smallest $\sigma_\mathrm{obs}$ measured in the main Fitts' law task was 1.591275 mm; the difference was only 0.001127 mm.
Because $\sigma_\mathrm{a}$ and $\sigma_\mathrm{obs}$ are computed from empirical data measured in a limited number of trials (i.e., random values) and assumed to be normally distributed, it is possible to observe $\sigma_\mathrm{a}$ greater than $\sigma_\mathrm{obs}$ by chance.
According to the assumption that the observed $\mathit{SD}$ increases as the target size increases, using a target with $W=1$ pixel would yield a smaller $\mathit{SD}$ than using $W=4.8$ mm for the calibration task.
Therefore, using a target with $W=1$ pixel for the calibration task should reduce the risk of having $\sigma_\mathrm{obs}^2 < \sigma_\mathrm{a}^2$ when analyzing the FFitts law fitness.

For this reason, using a 1-pixel target is more theoretically sound.
This solution was noticed by Bi et al., who used a 2.4-mm target for calibration (``Alternatively, single pixel wide lines and cross hairs could be used in lieu of bars and circles.'') \cite{Bi13a}.
Yet, a 1-pixel target is not an exact ``alternate'' for a 2.4-mm target (i.e., they are not \HL{interchangeable}), because the touch point variability should depend on the given target size according to Equation~\ref{eqn:sigma2_a2_W2} ($\sigma_\mathrm{obs}^2 = \alpha W^2 + \sigma_\mathrm{a}^2$).
More rigorously, Bi et al. hypothesized that $\sigma_\mathrm{a}$ would not be affected by the speed-accuracy rule, so $\sigma_\mathrm{a}$ must be the $\sigma_\mathrm{obs}$ value for the $W=0$ (either mm or pixels) condition.
Practically, however, the finest target must be visible; thus, $W=1$ pixel is a reasonable approximation of $W=0$ pixel.

\subsection{Instruction in Calibration Task}
\label{sec:calInst}
There have been two instruction choices: balancing the speed and accuracy \cite{Bi13a} or concentrating on accuracy \cite{Luo14,Yamanaka18mobilehci,Yamanaka18iss}.
We assume that both instructions are valid for measuring $\sigma_\mathrm{a}$.
For the ``rapid and accurate'' instruction by Bi et al., as the $W$ becomes smaller, participants have to be more careful to avoid missing the target, which causes them to spend a longer time.
Therefore, even if the participants were instructed to tap the target ``as rapidly (and accurately) as possible,'' the operational time for a 1-pixel (or smallest-$W$) target would be quite long, and the difference from the instruction to ``concentrate on accuracy'' would become almost negligible.
Still, the effect of this instruction difference on FFitts law fitness has been neither discussed nor empirically compared.
Hence, we empirically assess this difference in our data analyses.

\subsection{Computation of \texorpdfstring{$\sigma_\mathrm{a}$}{sigma a}: Calibration Task, Intercept of Regression, or Parameter Optimization}
\begin{figure}[t]
\begin{tabular}{lr}%fig and table = c, c

\hspace{-12pt} %minipageの合計を0.9くらいにして，左を微妙に左寄せ，右は微妙に右寄せにすると間隔をあけられる．
 \begin{minipage}{0.47\textwidth}
 \includegraphics[width=1.0\columnwidth]{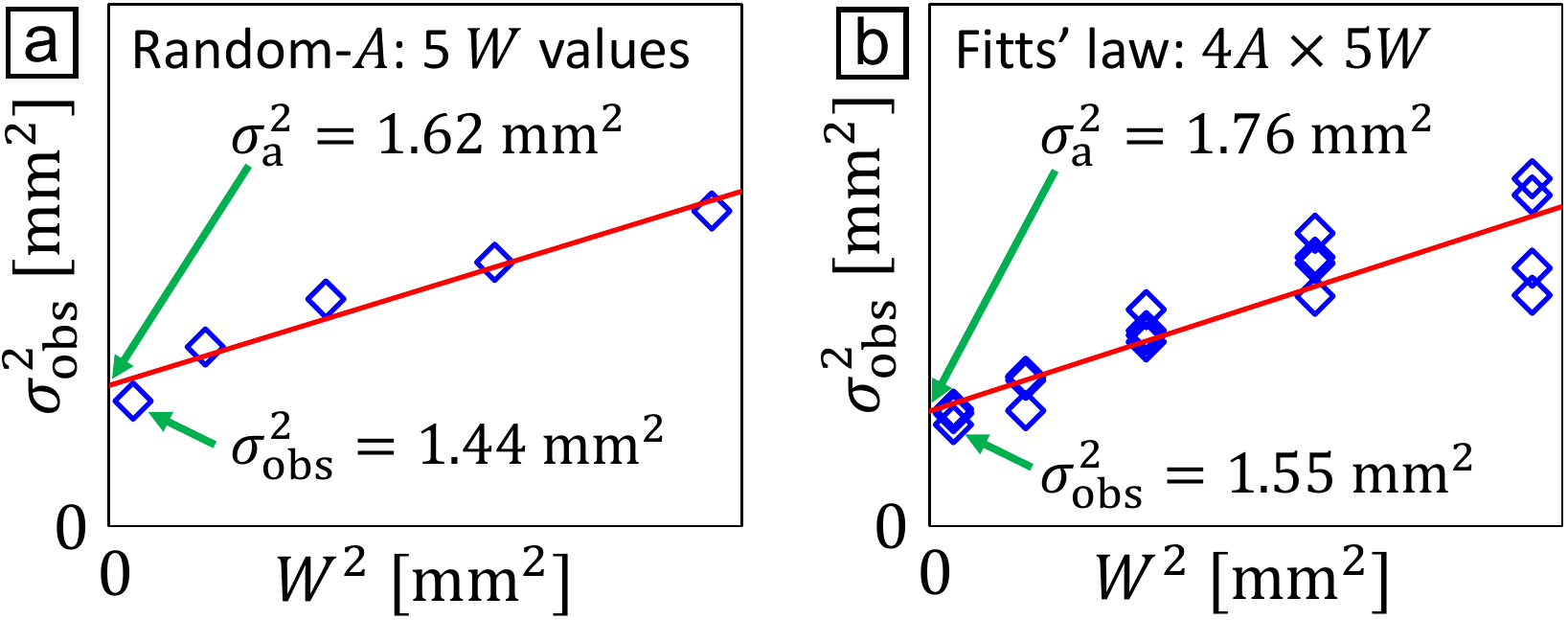}
 
 \caption{A problem in using the intercept method to compute $\sigma_\mathrm{a}$. The intercepts obtained from the (a) random-$A$ and (b) preset-$A$ tasks (i.e., Fitts' law) in our 2D experiment were greater than some of the $\sigma_\mathrm{obs}^2$ values in the Fitts' law task.}
 \label{fig:problems}
 \end{minipage}

\hspace{12pt} %%<- これを挿入値は好みにあわせる
 \begin{minipage}{0.45\textwidth}
 \raggedleft
 \includegraphics[width=1.0\columnwidth]{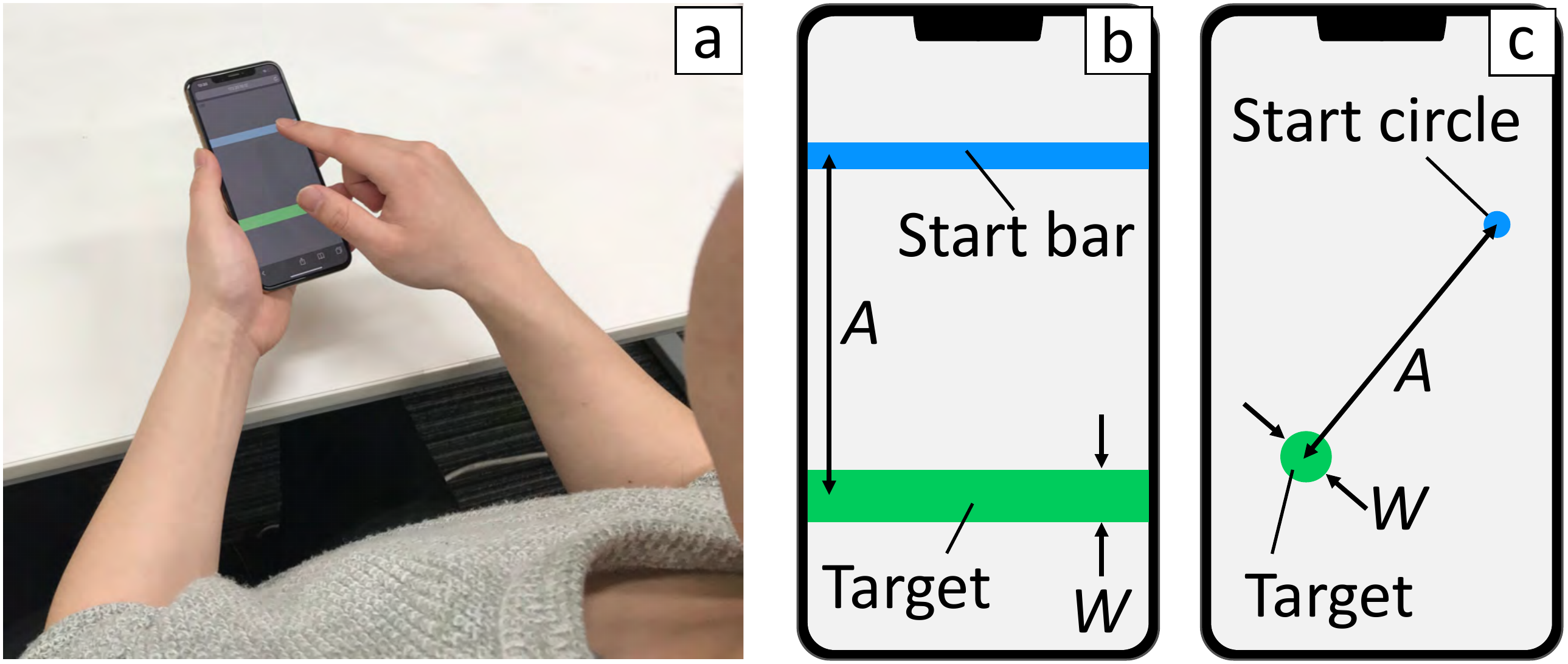}
 \caption{(a) A participant attempting a 1D Fitts' law task. The visual stimuli used in the (b) 1D and (c) 2D Fitts' law tasks.}
 \label{fig:studyScene}
 \end{minipage}
 
\end{tabular}
\end{figure}

To obtain $\sigma_\mathrm{a}$ by the intercept of the regression expression for $\sigma_\mathrm{obs}^2$ vs. $W^2$, Bi and Zhai \cite{Bi13b} and Yamanaka and Usuba \cite{Yamanaka20issFFF} conducted target-pointing tasks in which a new target appeared at a random position (i.e., $A$ was not controlled by the researchers), with several $W$ values; they then obtained regression expressions.
Yamanaka and Usuba also ran regressions for Fitts' law tasks in which four $A$ values were preset to use this method.
If we apply $\sigma_\mathrm{a}$ computed by this intercept method to FFitts law, it is possible to obtain $\sigma_\mathrm{a}$ greater than $\sigma_\mathrm{obs}$, which causes the mathematical error.
Figure~\ref{fig:problems} illustrates this problem with the data from our 2D experiment.
In this case, for the $\sigma_\mathrm{a}^2$ values computed from both the random- and preset-$A$ conditions, several $\sigma_\mathrm{obs}^2$ values at the lowest $W$ condition in the main Fitts' law task (Figure~\ref{fig:problems}b) are smaller than the intercept.

To avoid this issue, a possible choice is to use large $W$ values for the main Fitts' law task.
For example, if we had not used the narrowest $W$ condition in Figure~\ref{fig:problems}b, all the $\sigma_\mathrm{obs}$ values would be greater than $\sigma_\mathrm{a}$.
Using only wide $W$ values also lowers the risk of the mathematical error in using the $\sigma_\mathrm{a}$ measured by a finger calibration task.
Yet, this approach has a limitation: it prevents researchers from using a small target, and the threshold for the smallest target to avoid the error is unclear.
In addition, the effectiveness of FFitts law is for small targets; when targets are large, FFitts law approximates the original effective width method \cite{Bi13a}.

The state-of-the-art method to obtain the finger tremor factor is parameter optimization \cite{Ko20FF2D}.
The method's drawback is that it uses an additional free parameter $c$, which is adjusted to maximize $R^2$ for the regression of $\mathit{MT}$ vs. $\mathit{ID}$.
Generally speaking, introducing additional free parameters could lead to overfitting. 
In contrast, using a $\sigma_\mathrm{a}$ value computed from a calibration task or the intercept method has no such problem, because $\sigma_\mathrm{a}$ is independent of the $\mathit{MT}$ values measured in a Fitts' law task.

Regarding the model fitness in terms of $R^2$, using parameter optimization would theoretically give the best fit among the candidates.
Also, it does not require an independent finger calibration task and is thus less time-consuming for researchers and participants.
However, if other model-fit metrics that consider the model complexity show a worse result due to the free parameter $c$, then using $\sigma_\mathrm{a}$ instead of $c$ is recommended.
To assess this issue, we also compare the model fitness by using the adjusted $R^2$, Akaike Information Criterion $\mathit{AIC}$, Bayesian Information Criterion $\mathit{BIC}$, and root-mean-square error $\mathit{RMSE}$ of the cross-validation in our data analyses.

\section{Experiments}
We conducted touch-pointing experiments with a smartphone, as shown in Figure~\ref{fig:studyScene}a.
The experiments were conducted on two separate days: Day 1 for 1D horizontal bar-shaped targets, and Day 2 for 2D circular targets.
The procedures for the two days were the same.
Under both the 1D and 2D conditions, we conducted four sub-tasks.
The main one was a Fitts' law task with $4A \times 5W$ conditions, and the remaining three sub-tasks were used to compute $\sigma_\mathrm{a}$ values: one was for the intercept-based method with five $W$ values and random $A$ values, and the other two were for finger calibration tasks.
The order of the four sub-tasks was balanced using a Latin square pattern among 12 participants for both days.
Each participant took 40 to 50 min for the experiment on each day.

For both the 1D and 2D conditions, our $\sigma_\mathrm{a}$ data computed by the intercept method for Fitts' law and the random-$A$ tasks was reported before \cite{Yamanaka20issFFF}.
The data for the two finger calibration tasks is newly reported here.
Because our novel contribution in this paper is the evaluation of the model fitness for $\mathit{MT}$, we repeat the minimum necessary explanation of the experiments (e.g., the mean error rate) to make this paper self-contained, while taking care to avoid plagiarism.
For example, we could have reported all the pairwise test results for the error rate, but that data would not relate to this paper's main contribution.
Thus, we mainly report the $\mathit{MT}$ and $\sigma_\mathrm{obs}$ results, and readers who are interested in the detailed error-rate prediction models are directed to \cite{Yamanaka20issFFF}.

\subsection{Sub-Tasks}
\subsubsection{Finger Calibration Task with ``Rapid and Accurate'' Instruction}
\label{sec:1DcalibQandA}
The participants were instructed to tap as rapidly and accurately as possible on a 1-pixel horizontal bar target or a 25-pixel-wide crosshair target in the 1D or 2D conditions, respectively.
For the 2D condition, we emphasized that the intersection of the crosshair was the target to aim for.
A 1-sec break was enforced before the next target appeared as in \cite{Bi13a,Woodward20avi}.
Each participant repeated this procedure 50 times, which entailed five practice trials followed by 45 data-collection trials.
The signed biases of the tap point from the target were used to compute the $\mathit{SD}$ (i.e., $\sigma_\mathrm{a}$) on the y-axis for the 1D case and the bivariate $\mathit{SD}$ on the x- and y-axes for the 2D case.

\subsubsection{Finger Calibration Task with ``Concentrate on Accuracy'' Instruction}
For this sub-task, only the instruction was different from the previously explained sub-task.
That is, the participants were instructed to tap as closely as possible to the target without paying attention to the operational time.

\subsubsection{Fitts' Law Task}
This was a discrete pointing task with preset $A$ and $W$ values.
For the 1D task, a 6-mm-wide blue start bar was displayed at the top of the screen, and a green target bar was at the bottom, as shown in Figure~\ref{fig:studyScene}b.
The movement direction was always downwards.
When participants tapped the start bar, it disappeared, and a click sound played.
Then, if they successfully tapped the target, a pleasant bell played, and then the next set of start and target bars appeared.
If the tap point fell outside the target, they had to aim for the target again until they succeeded; the trial was not restarted from tapping the start bar.
The participants were instructed to tap the target as rapidly and accurately as possible.
For the 2D task, circles were used instead of horizontal bars, and the start and target circles' positions were randomized while keeping a distance $A$ between them.

This sub-task used a $4 \times 5$ within-subjects design with the following independent variables and levels.
We included four target distances ($A$ = 20, 30, 45, and 60 mm) and five target widths ($W$ = 2, 4, 6, 8, 10 mm).
Each $A \times W$ combination entailed a single repetition of practice trials followed by 16 repetitions.
The order of the $20$ conditions was randomized.
Thus, we recorded $4_A \times 5_W \times 16_\mathrm{repetitions} \times 12_\mathrm{participants} = 3840$ data points in total.
The dependent variables were the $\mathit{MT}$, the standard deviation of the endpoints ($\sigma_\mathrm{obs}$), and the error rate.

\subsubsection{Pointing Task with Random Target Distance}
For the 1D case, a 6-mm-high start bar was initially displayed at a random position.
When the participants tapped it, the first target bar appeared at a random position, and then they successively tapped new targets.
If a target was missed, a beep sounded, and the participants re-aimed for the target.
A successful tap resulted in a bell sound.
For the 2D case, circular targets were used.

This sub-task used a single-factor, within-subjects design with an independent variable of $W$: 2, 4, 6, 8, and 10 mm.
The dependent variable was the observed touch-point distribution, $\sigma_\mathrm{obs}$.
First, the participants performed 20 trials as practice, which included 4 repetitions of the 5 $W$ values appearing in random order.
In each \textit{session}, the $W$ values appeared 10 times in a random order.
The participants were instructed to successively tap the target as rapidly and accurately as possible in a session.
They each completed four sessions as data-collection trials.
In total, we recorded $5_W \times 10_\mathrm{repetitions} \times 4_\mathrm{sessions} \times 12_\mathrm{participants} = 2400$ trials.

\subsection{Participants}
On Day 1, 12 university students participated in this study (2 female, 10 male; 20 to 25 years, $M=23.0$, $\mathit{SD}=1.41$).
On Day 2, 12 university students again participated (3 female, 9 male; 19 to 25 years, $M=22.2$, $\mathit{SD}=2.12$), with nine new participants.
For both days, all the participants had normal or corrected-to-normal vision.
All were right-handed and were daily smartphone users.
Each participant received JPY 5000 ($\sim$US\$ 45) in compensation for one day.
The participants were instructed to hold the smartphone in their non-dominant (left) hand and perform tapping operations with their dominant (right) index finger.

They were instructed to sit on an office chair, but to increase the generalizability of our findings, they were asked not to rest their hands or elbows on the table or their lap.
This decision was made because stabilizing the hand with the support of the elbow significantly improves the touch precision \cite{Ikematsu20iss}, but in our daily life, using a smartphone while standing or walking (i.e., \HL{with no} stabilization) is also common.
Still, to reduce the negative effects of fatigue on the results, we instructed the participants to take a break if needed when operational data was not measured.

\subsection{Apparatus}
On both days, we used an iPhone XS Max (4 GB RAM; iOS 12; $1242 \times 2688$ pixels, 6.5-inch-diagonal display, 458 ppi).
We implemented the experimental system as a webpage by using JavaScript, HTML, and CSS.
The Safari app was used to view the webpage.
After eliminating the navigation-bar areas, the canvas resolution was converted to $414 \times 719$ pixels (5.978 pixels/mm resolution).
The system was set to run at 60 fps.
We used the take-off positions as tap points, as in previous studies \cite{Bi13a,Bi13b,Bi16,Yamanaka18mobilehci,Yamanaka18iss}.

Note that we also examined the results using the land-on positions and $\mathit{MT}$s.
The mean absolute differences between the land-on and take-off positions were less than 1 pixel on both the x- and y-axes in all eight sub-tasks, which was smaller than the minimum value that our apparatus could sense.
In addition, while the $\mathit{MT}$s for the land-on timing were approximately 80~ms earlier than those for the take-off timing, this only affected the intercept $a$ in the models and did not affect the model fitness.
Therefore, for consistency with previous studies, we report here only the results using the take-off positions and timings.

\section{Results of 1D Experiment}
As in previous studies, data points for which the distance between the tap point and the target center was greater than 15 mm were removed as outliers before we analyzed the $MT$, $\sigma_\mathrm{a}$, $\sigma_\mathrm{obs}$, and error rate \cite{Bi13b,Yamanaka20issFFF}.
Using a fixed distance may affect $W$ levels differently, e.g., more outliers may be observed for $W = 10$ mm than $W = 2$ mm, but in this paper, we maintain consistency with the previous study \cite{Bi13b}.

\subsection{Finger Calibration Task with ``Rapid and Accurate'' Instruction}
Among the 540 trials (45 repetitions $\times$ 12 participants), we observed no outliers.
Two participants' data did not pass the normality test (Shapiro-Wilk test with alpha $= 0.05$).
The $\mathit{SD}$ of the tap positions (i.e., $\sigma_\mathrm{a}$) for each participant ranged from 0.5448 to 1.325 mm, and the mean was 0.8837 mm.

\subsection{Finger Calibration Task with ``Concentrate on Accuracy'' Instruction}
We again observed no outliers, while two participants' data did not pass the normality test.
The $\sigma_\mathrm{a}$ values ranged from 0.4569 to 1.296 mm among the participants, and the mean $\sigma_\mathrm{a}$ was 0.7362 mm.

\subsection{Fitts' Law Task}
Among the 3840 trials, four data points were removed as outliers (0.10\%).
The outliers resulted mainly from participants accidentally touching the screen with the thumb or little finger.
Two or more taps were observed in 347 trials, and the mean error rate was thus 9.046\%.
We found that 218 of the 240 conditions ($4_A \times 5_W \times 12_\mathrm{participants}$) passed the normality test, or 90.8\%.

We use RM-ANOVA with Bonferroni's $p$-value adjustment method for pairwise comparisons, because it is known that ANOVA is robust against violations of the normality test assumptions for dependent variables \cite{dixon2008models,mena2017non}.
For the $F$ statistic, the degrees of freedom for the main effects of $A$ and $W$, as well as their interactions, were corrected using the Greenhouse-Geisser method when Mauchly's sphericity assumption was violated.

For the endpoint variability $\sigma_\mathrm{obs}$, we found significant main effects of $A$ ($F_{3,33} = 2.949$, $p<0.05$, {\small $\eta_p^2$}$= 0.21$) and $W$ ($F_{4,44} = 72.63$, $p<0.001$, {\small $\eta_p^2$}$= 0.87$), but no significant interaction of $A \times W$ ($F_{12,132} = 1.371$, $p=0.187$, {\small $\eta_p^2$}$= 0.11$).
For the $\mathit{MT}$, we found significant main effects of $A$ ($F_{3,33} = 201.499$, $p<0.001$, {\small $\eta_p^2$}$= 0.95$) and $W$ ($F_{4,44} = 89.699$, $p<0.001$, {\small $\eta_p^2$}$= 0.89$), and the interaction of $A \times W$ was significant ($F_{12,132} = 4.417$, $p<0.001$, {\small $\eta_p^2$}$= 0.29$).

Figure~\ref{fig:reg1D_sigma_w}a shows the result of $\sigma_\mathrm{obs}^2$ vs. $W^2$ regression.
The $\sigma_\mathrm{a}$ value was $\sqrt{0.9543} = 0.9769$ mm.
The regression line clearly passes above the four data points at the smallest $W^2$ value.
Importantly, in this case, the intercept $\sigma_\mathrm{a}^2$ was greater than some $\sigma_\mathrm{obs}^2$, causing the mathematical error in FFitts law.

\subsection{Pointing Task with Random Target Distance}
\begin{figure}[t]
\centering
\includegraphics[width=0.67\columnwidth]{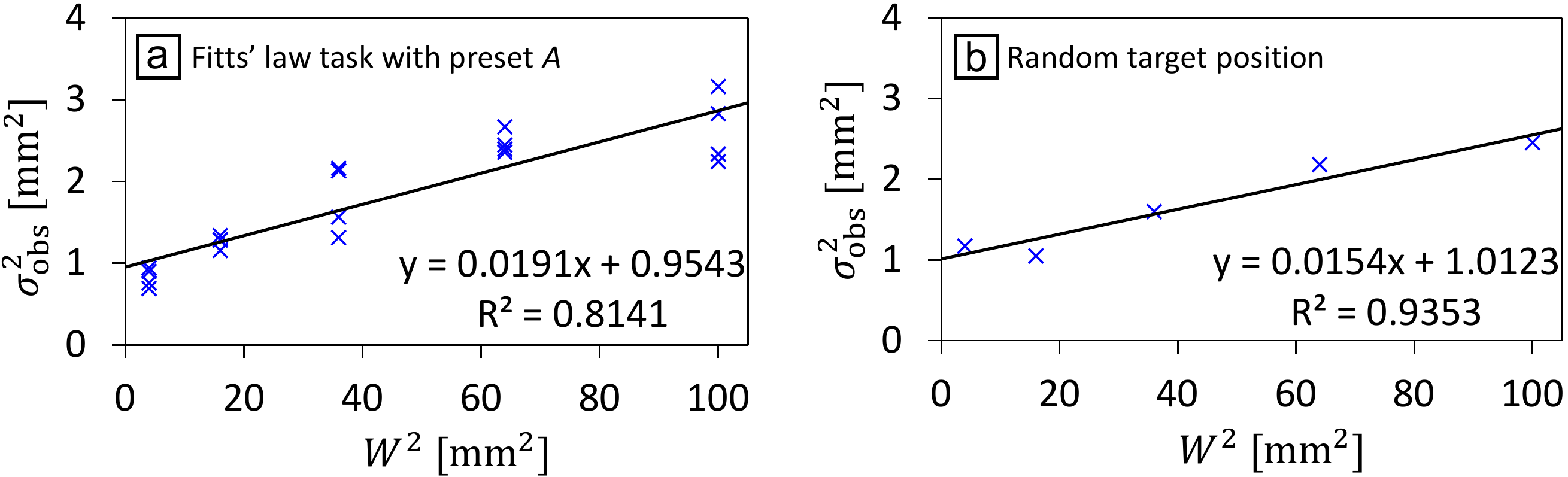}
\caption{Regression results of $\sigma_\mathrm{obs}^2$ vs. $W^2$ for the 1D conditions. The intercepts show the $\sigma_\mathrm{a}^2$ values.}
\label{fig:reg1D_sigma_w}
\end{figure}

We removed 13 outlier trials (0.54\%).
The Shapiro-Wilk test showed that the touch points followed a normal distribution under 47 of the 60 conditions ($ = 5_W \times 12_\mathrm{participants}$), or 78.3\%.
RM-ANOVA showed that $W$ significantly affected $\sigma_\mathrm{obs}$ ($F_{4,44} = 11.18$, $p<0.001$, {\small $\eta_p^2$}$= 0.50$).
Figure~\ref{fig:reg1D_sigma_w}b shows the regression result.
The value of $\sigma_\mathrm{a}$ was $\sqrt{1.0123} = 1.006$ mm, and some $\sigma_\mathrm{obs}^2$ values in the Fitts' law task (Figure~\ref{fig:reg1D_sigma_w}a) were smaller than 1.0123 mm, which caused the mathematical error when we applied the $\sigma_\mathrm{a}$ measured with this intercept method to FFitts law data.

After obtaining $\sigma_\mathrm{a}$ values by the four methods (i.e., sub-tasks), we ran a non-parametric ANOVA with \HL{the} aligned rank transform \cite{Wobbrock11ART}.
The result showed that the computation method significantly affected $\sigma_\mathrm{a}$ ($F_{3,33} = 8.266$, $p<0.001$, {\small $\eta_p^2$}$= 0.43$).
Pairwise tests with Tukey's $p$-value adjustment showed that the $\sigma_\mathrm{a}$ for \HL{the} ``Concentrate on Accuracy'' calibration task was smaller than that for \HL{the} Fitts' law task ($p < 0.001$) and \HL{the} random-$A$ task ($p < 0.001$).

\subsection{Model Fitting Results for 1D Task}
We found an issue when analyzing the FFitts law fitness: the finger calibration and intercept methods could not be used because of the mathematical error, and thus, we could only use the parameter optimization method of FFitts law.
As listed in Table~\ref{tab:1D_W_f}, among the $4_A \times 5_W = 20$ data points for fitting, any method using $\sigma_\mathrm{a}$ had one or more mathematical errors (due to a negative value inside the square root in $W_f=\sqrt{2 \pi e (\sigma_\mathrm{obs}^2 - \sigma_\mathrm{a}^2)}$).
This result shows the low robustness of FFitts law when using $\sigma_\mathrm{a}$, regardless of whether the $\sigma_\mathrm{a}$ value is directly measured by a finger calibration task or calculated by the intercept method.

For model fitness comparison, we use the absolute and adjusted $R^2$.
The latter balances the number of coefficients.
We also compare models through the $\mathit{AIC}$ \cite{Akaike74}.
This statistical method balances the number of free parameters and the fitness to identify a comparatively best model.
As a brief guideline, (a) a model with a lower $\mathit{AIC}$ value is a better one; (b) a model with $\mathit{AIC}$ $\leq$ ($AIC_\mathrm{minimum} + 2$) is probably comparable with better models; and (c) a model with $\mathit{AIC}$ $\geq$ ($AIC_\mathrm{minimum} + 10$) should be rejected.
We also use the $\mathit{BIC}$ \cite{Robert95bic} for comparison: $\mathit{BIC}$ differences of 0--2 are not significant, of 2--6 are positive, and of 6--10 are strong; differences greater than 10 are very strong \cite{Robert95bic}.
The $\mathit{AIC}$ penalizes \HL{the use of} additional free parameters the least, while the $\mathit{BIC}$ penalizes it the most.

Moreover, we ran a leave-one-($A, W$)-out cross-validation and computed the $\mathit{RMSE}$.
Similar to the $\mathit{AIC}$ and $\mathit{BIC}$, \HL{the} cross-validation is beneficial \HL{for judging} whether it is worth introducing extra free parameter(s) and checking whether there is an overfitting issue.
Also, if the fitness values in terms of \HL{the} adjusted $R^2$ or information criteria are close to each other for several models, \HL{the} cross-validation could provide extra information to determine a better model.
Because our purpose \HL{in} this study is to compare the prediction accuracy of $MT$, \HL{it is helpful that the cross-validation has predictive power for unknown (new) task conditions.}
A model with a higher $R^2$ and adjusted $R^2$ is better, while one with a lower $\mathit{AIC}$, $\mathit{BIC}$, and $\mathit{RMSE}$ is also better.

\begin{figure}[t]
\centering
\includegraphics[width=0.95\columnwidth]{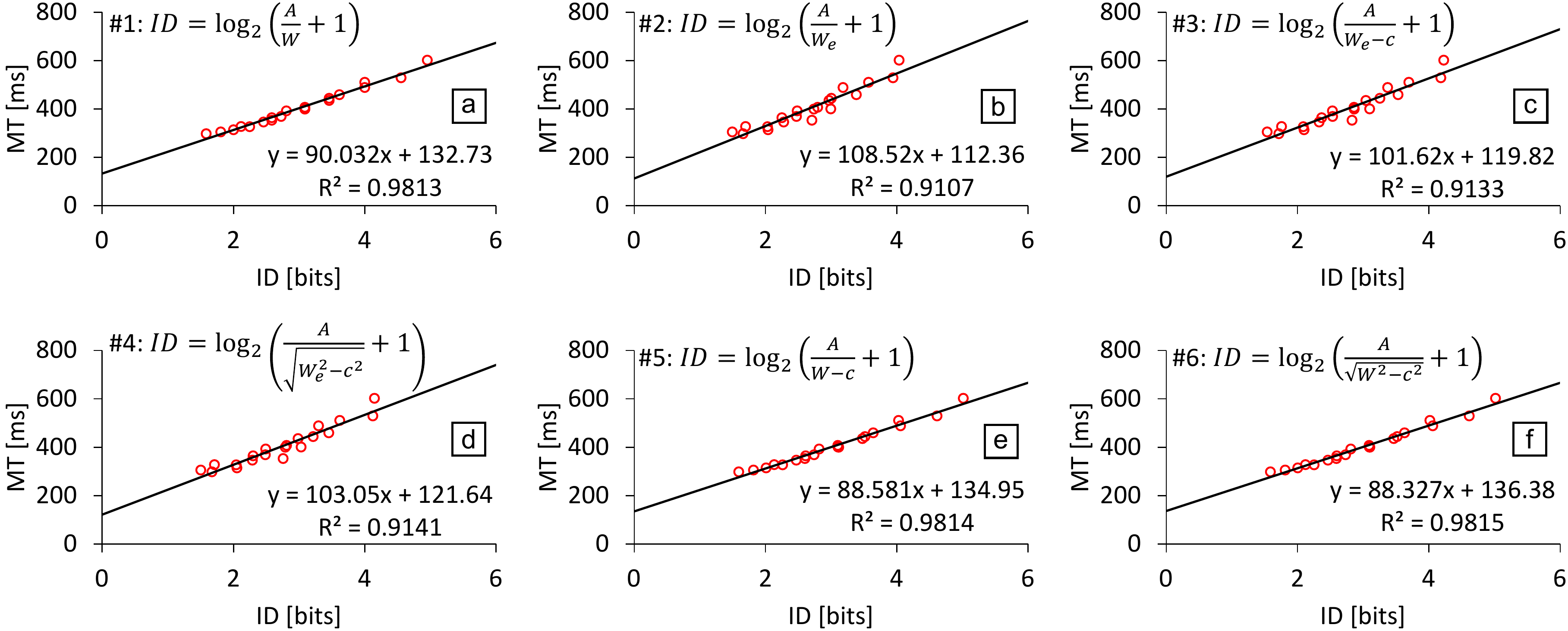}
\caption{$\mathit{MT}$ vs. $\mathit{ID}$ regressions of Models \#1 to \#6 in Table~\ref{tab:1D_fit} for the 1D tasks.}
\label{fig:e1fitGraph}
\end{figure}

Table~\ref{tab:1D_fit} lists the model fitness results.
Overall, the baseline model of Fitts' law showed the best model fitness in terms of the adjusted $R^2$, $\mathit{AIC}$, $\mathit{BIC}$, and $\mathit{RMSE}$.
While Model \#6 showed the highest $R^2$, it was due to the additional free parameter; thus, the adjusted $R^2$ was slightly lower than that of Model \#1.
According to the $\mathit{AIC}$ and $\mathit{BIC}$, Models \#5 and \#6 should not be rejected as worse models than Model \#1.
Lastly, the models using $W_e$ (\#2, \#3, and \#4) are significantly worse than the other models and can be safely rejected.
These lower fits are also shown in Figure~\ref{fig:e1fitGraph}.
As a result, we empirically confirmed that the baseline Model \#1 is the best, and \HL{that the introduction of another free parameter $c$ (Models \#5 and \#6) yields comparable fits.}

\renewcommand{\arraystretch}{1.2}%height of table row
\begin{table}[t]
\caption{Measured data for the 1D tasks. The units are all in mm, except for the $\mathit{MT}$ in ms. The four $\sigma_\mathrm{a}$ values were computed from the data as follows. Calib (R\&A): the finger calibration task with the ``rapid and accurate'' instruction. Calib (Acc): the finger calibration task with the ``concentrate on accuracy'' instruction. Fitts: the intercept method for the Fitts' law task. Random $A$: the intercept method for the pointing task with a random target position. The $W_f$ values were calculated by Equation~\ref{eqn:Wf_Bi} ($W_f=\sqrt{2 \pi e (\sigma_\mathrm{obs}^2 - \sigma_\mathrm{a}^2)}$). \HL{In the yellow cells, ``!err'' indicates the mathematical error.}}
\label{tab:1D_W_f}
\scalebox{0.675}{%use small table
\begin{tabular}{l | c || c || r|r|r|r | r|r|r|r | r|r|r|r | r|r|r|r | r|r|r|r }
\hline

& & $A$ & 20 & 20 & 20 & 20 & 20 & 30 & 30 & 30 & 30 & 30 & 45 & 45 & 45 & 45 & 45 & 60 & 60 & 60 & 60 & 60 \\
& & $W$ & 2 & 4 & 6 & 8 & 10 & 2 & 4 & 6 & 8 & 10 & 2 & 4 & 6 & 8 & 10 & 2 & 4 & 6 & 8 & 10 \\ \wcline{1-23}
& & $\mathit{MT}$ & 444 & 364 & 328 & 305 & 298 & 489 & 400 & 353 & 327 & 315 & 529 & 459 & 400 & 369 & 347 & 602 & 511 & 436 & 407 & 393 \\
& $\sigma_\mathrm{a}$ & $\sigma_\mathrm{obs}$ & 0.69 & 1.29 & 2.16 & 2.66 & 2.24 & 0.899 & 1.28 & 1.31 & 2.36 & 2.33 & 0.757 & 1.16 & 1.56 & 2.39 & 2.83 & 0.942 & 1.34 & 2.13 & 2.44 & 3.16 \\ \cline{1-23}
Calib (R\&A) & 0.884 & $W_f$ & \cellcolor[rgb]{1,1,0.6}{!err} & 3.87 & 8.15 & 10.4 & 8.52 & 0.694 & 3.83 & 4.01 & 9.02 & 8.93 & \cellcolor[rgb]{1,1,0.6}{!err} & 3.07 & 5.32 & 9.20 & 11.1 & 1.35 & 4.14 & 8.01 & 9.41 & 12.5 \\
Calib (Acc) & 0.736 & $W_f$ & \cellcolor[rgb]{1,1,0.6}{!err} & 4.36 & 8.39 & 10.6 & 8.75 & 2.14 & 4.33 & 4.49 & 9.25 & 9.15 & 0.728 & 3.68 & 5.69 & 9.42 & 11.3 & 2.43 & 4.61 & 8.26 & 9.63 & 12.7 \\
Fitts & 0.977 & $W_f$ & \cellcolor[rgb]{1,1,0.6}{!err} & 3.47 & 7.96 & 10.2 & 8.34 & \cellcolor[rgb]{1,1,0.6}{!err} & 3.42 & 3.62 & 8.86 & 8.76 & \cellcolor[rgb]{1,1,0.6}{!err} & 2.55 & 5.04 & 9.04 & 11.0 & \cellcolor[rgb]{1,1,0.6}{!err} & 3.77 & 7.82 & 9.26 & 12.4 \\
Random $A$ & 1.01 & $W_f$ & \cellcolor[rgb]{1,1,0.6}{!err} & 3.32 & 7.90 & 10.2 & 8.28 & \cellcolor[rgb]{1,1,0.6}{!err} & 3.27 & 3.48 & 8.8 & 8.70 & \cellcolor[rgb]{1,1,0.6}{!err} & 2.34 & 4.94 & 8.98 & 10.9 & \cellcolor[rgb]{1,1,0.6}{!err} & 3.64 & 7.76 & 9.20 & 12.4 \\

\end{tabular}
}%end small
\bigskip\centering
\end{table}
\renewcommand{\arraystretch}{1.0}%height of table row

\renewcommand{\arraystretch}{1.5}%height of table row
\begin{table}[t]
\caption{Model fitness results for the 1D tasks. The yellow cells indicate the best fit for each criterion. The light-blue cells indicate the candidate best-fit models (i.e., those whose $\mathit{AIC}$ and $\mathit{BIC}$ differences from the best model are less than 10).}
\label{tab:1D_fit}
\scalebox{0.85}{%use small table
\begin{tabular}{l | l || c|c|c|c|c|| c|c|c}
\hline

\multicolumn{1}{c|}{Description} & \multicolumn{1}{c||}{$\mathit{ID}$ formulation} & $R^2$ & adj. $R^2$ & $\mathit{AIC}$ & $\mathit{BIC}$ & $\mathit{RMSE}$ & $a$ & $b$ & $c$ \\ \wcline{1-10}
\#1 Baseline & $\log_{2}\left(A/W+1\right)$ & 0.9813 & \cellcolor[rgb]{1,1,0.6}{0.9802} & \cellcolor[rgb]{1,1,0.6}{156.6} & \cellcolor[rgb]{1,1,0.6}{158.6} & \cellcolor[rgb]{1,1,0.6}{13.30} & 132.7 & 90.03 & --- \\ \hline
\#2 $\mathit{ID}_e$ & $\log_{2}\left(A/W_e+1\right)$ & 0.9107 & 0.9058 & 187.8 & 189.8 & 27.64 & 112.4 & 108.5 & --- \\ \hline
\#3 Param. Opt. ($W_e$, no sqrt) & $\log_2 \left( \frac{A}{ W_e - c} +1 \right)$ & 0.9133 & 0.9031 & 189.2 & 192.2 & 27.92 & 119.8 & 101.6 & 0.5067 \\ \hline
\#4 Param. Opt. ($W_e$, sqrt) & $\log_2 \left( \frac{A}{ \sqrt{ W_e^2 - c^2}} +1 \right)$ & 0.9141 & 0.9040 & 189.1 & 192.0 & 28.08 & 121.6 & 103.0 & 1.512 \\ \hline
\#5 Param. Opt. ($W$, no sqrt) & $\log_2 \left( \frac{A}{ W - c} +1 \right)$ & 0.9814 & 0.9792 & \cellcolor[rgb]{0.7,1,1}{158.5} & \cellcolor[rgb]{0.7,1,1}{161.5} & 13.76 & 134.9 & 88.58 & 0.08178 \\ \hline
\#6 Param. Opt. ($W$, sqrt) & $\log_2 \left( \frac{A}{ \sqrt{ W^2 - c^2}} +1 \right)$ & \cellcolor[rgb]{1,1,0.6}{0.9815} & 0.9793 & \cellcolor[rgb]{0.7,1,1}{158.3} & \cellcolor[rgb]{0.7,1,1}{161.3} & 13.96 & 136.4 & 88.33 & 0.5806 \\ \hline
 
\end{tabular}
}%end small
\bigskip\centering
\end{table}
\renewcommand{\arraystretch}{1.0}%height of table row

\section{Results of 2D Experiment}
\subsection{Finger Calibration Task with ``Rapid and Accurate'' Instruction}
Again, data points for which the distance between the tap point and the target center was longer than 15 mm were removed as outliers.
Among the 540 trials for this sub-task, we observed no outliers.
Two participants' data did not pass the normality test.
The $\mathit{SD}$ of the tap positions (i.e., $\sigma_\mathrm{a}$) for each participant ranged from 0.8717 to 2.148 mm, and the mean was 1.372 mm.

\subsection{Finger Calibration Task with ``Concentrate on Accuracy'' Instruction}
We again observed no outliers, while three participants' data did not pass the normality test.
The $\sigma_\mathrm{a}$ values ranged from 0.7107 to 1.752 mm among the participants, and the mean $\sigma_\mathrm{a}$ was 1.163 mm.

\subsection{Fitts' Law Task}
Among the 3840 trials, nine outlier trials were removed (0.23\%).
The mean error rate was 17.91\%.
Under 184 (76.7\%) conditions, the touch points followed a bivariate normal distribution.
For the tap point distribution $\sigma_\mathrm{obs}$, we found a significant main effect of $W$ ($F_{4,44} = 47.606$, $p<0.001$, {\small $\eta_p^2$}$= 0.82$), but not of $A$ ($F_{3,33} = 2.787$, $p=0.056$, {\small $\eta_p^2$}$= 0.20$).
The interaction of $A \times W$ was also not significant ($F_{12,132} = 1.151$, $p=0.325$, {\small $\eta_p^2$}$= 0.10$).
For the $\mathit{MT}$, we found significant main effects of $A$ ($F_{3,33} = 181.376$, $p<0.001$, {\small $\eta_p^2$}$= 0.94$) and $W$ ($F_{1.114,12.256} = 69.498$, $p<0.001$, {\small $\eta_p^2$}$= 0.86$), and the interaction of $A \times W$ was significant ($F_{4.636,50.991} = 6.450$, $p<0.001$, {\small $\eta_p^2$}$= 0.33$).

Figure~\ref{fig:reg2D_sigma_w}a shows the result of $\sigma_\mathrm{obs}^2$ vs. $W^2$ regression.
The $\sigma_\mathrm{a}$ value was $\sqrt{1.7593} = 1.326$ mm.
The regression line passes above the four data points at the smallest $W^2$ value, causing the mathematical error in FFitts law.

\subsection{Pointing Task with Random Target Distance}
\begin{figure}[t]
\centering
\includegraphics[width=0.72\columnwidth]{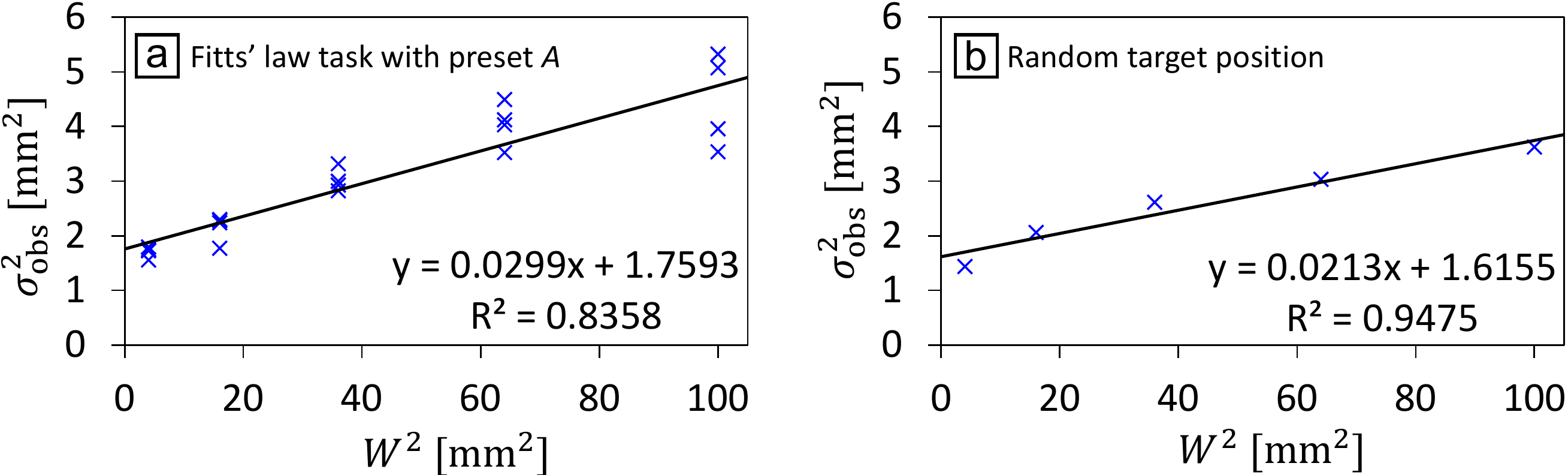}
\caption{Regression results of $\sigma_\mathrm{obs_y}^2$ vs. $W^2$ for 2D conditions. The intercepts show the $\sigma_\mathrm{a}^2$ values.}
\label{fig:reg2D_sigma_w}
\end{figure}

We removed 33 outlier trials (1.375\%).
Under 41 (68.3\%) conditions, the touch points followed a bivariate normal distribution.
The value of $W$ had a significant main effect on $\sigma_\mathrm{obs}$ ($F_{4,44} = 34.794$, $p<0.001$, {\small $\eta_p^2$}$= 0.76$).
Figure~\ref{fig:reg2D_sigma_w}b shows the regression result.
The value of $\sigma_\mathrm{a}$ was $\sqrt{1.6155} = 1.271$ mm, and this $\sigma_\mathrm{a}^2$ was greater than some $\sigma_\mathrm{obs}^2$ values in the Fitts' law task (Figure~\ref{fig:reg2D_sigma_w}a), which caused the mathematical error.
A non-parametric ANOVA with \HL{the} aligned rank transform showed that the computation method \HL{did not significantly affect} $\sigma_\mathrm{a}$ ($F_{3,33} = 1.738$, $p = 0.1784$, {\small $\eta_p^2$}$= 0.14$).

\subsection{Model Fitting Results for 2D Task}
\begin{figure}[t]
\centering
\includegraphics[width=0.99\columnwidth]{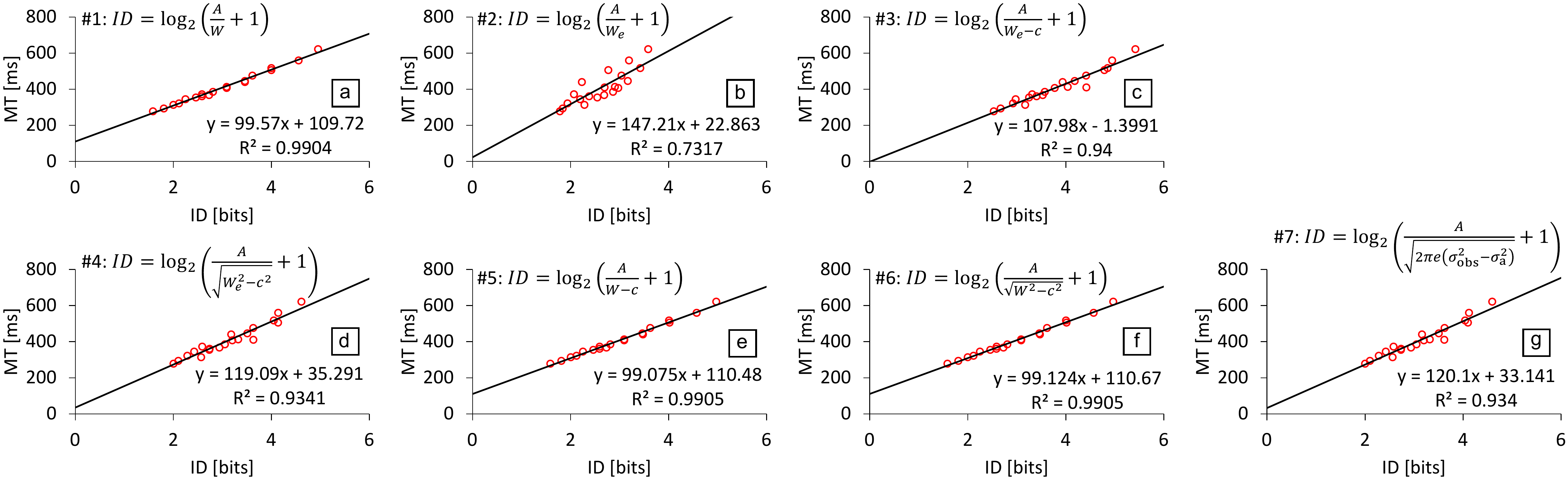}
\caption{$\mathit{MT}$ vs. $\mathit{ID}$ regressions of Models \#1 to \#7 in Table~\ref{tab:2D_fit} for the 2D tasks.}
\label{fig:e2fitGraph}
\end{figure}

In contrast to the results for the 1D task, we can use the $\sigma_\mathrm{a}$ value obtained from the finger calibration task with the ``concentrate on accuracy'' instruction.
In this case, as listed in Table~\ref{tab:2D_W_f}, the $4_A \times 5_W = 20$ data points for fitting had no negative values inside the square root in FFitts law.
Thus, in Table~\ref{tab:2D_fit}, we add Model \#7, which is the original FFitts law model.
The fitting results are also shown in Figure~\ref{fig:e2fitGraph}.

Overall, for Models \#1 to \#6, the results were similar to those for the 1D tasks.
The baseline Model \#1 was the best in terms of the adjusted $R^2$, $\mathit{AIC}$, $\mathit{BIC}$, and $\mathit{RMSE}$ values.
According to the $\mathit{AIC}$ and $\mathit{BIC}$, Models \#5 and \#6 showed similar model fitness to that of Model \#1.
The $\mathit{RMSE}$ value for Model \#6 \HL{was} close to that of \#1 (11 and 10 ms, respectively), but that for \#5 \HL{was} larger (42 ms).
The models using $W_e$ (\#2, \#3, and \#4) \HL{were} significantly worse.
Regarding Model \#7, which also uses the $W_e$ factor ($2 \pi e \sigma_\mathrm{obs}^2=W_e^2$), it showed significantly worse fits than those of Models \#1, \#5, and \#6, but the fit was improved \HL{in comparison} with the original effective width method (\#2).
The conclusion obtained from the 2D tasks is equivalent to that obtained from the 1D tasks: we empirically confirmed that the baseline model is the best, and \HL{the introduction of} a free parameter $c$ (Models \#5 and \#6) \HL{yields} comparable fits, while \#6 is better than \#5 according to the cross-validation.

\renewcommand{\arraystretch}{1.2}%height of table row
\begin{table}[t]
\caption{Measured data for the 2D tasks. Calib (R\&A): the finger calibration task with the ``rapid and accurate'' instruction. Calib (Acc): the finger calibration task with the ``concentrate on accuracy'' instruction. Fitts: the intercept method for the Fitts' law task. Random $A$: the intercept method for the pointing task with a random target position. The $W_f$ values were calculated by Equation~\ref{eqn:Wf_Bi}. \HL{In the yellow cells, ``!err'' indicates the mathematical error.}}
\label{tab:2D_W_f}
\scalebox{0.685}{%use small table
\begin{tabular}{l | c || c || r|r|r|r | r|r|r|r | r|r|r|r | r|r|r|r | r|r|r|r }
\hline

& & $A$ & 20 & 20 & 20 & 20 & 20 & 30 & 30 & 30 & 30 & 30 & 45 & 45 & 45 & 45 & 45 & 60 & 60 & 60 & 60 & 60 \\
& & $W$ & 2 & 4 & 6 & 8 & 10 & 2 & 4 & 6 & 8 & 10 & 2 & 4 & 6 & 8 & 10 & 2 & 4 & 6 & 8 & 10 \\ \wcline{1-23}
& & $\mathit{MT}$ & 440 & 373 & 322 & 294 & 278 & 506 & 410 & 361 & 345 & 314 & 560 & 476 & 413 & 368 & 354 & 622 & 517 & 446 & 407 & 385 \\
& $\sigma_\mathrm{a}$ & $\sigma_\mathrm{obs}$ & 1.31 & 1.51 & 1.71 & 1.88 & 1.99 & 1.25 & 1.33 & 1.73 & 2.03 & 1.88 & 1.34 & 1.51 & 1.68 & 2.01 & 2.25 & 1.32 & 1.49 & 1.82 & 2.12 & 2.31 \\ \hline
Calib (R\&A) & 1.37 & $W_f$ & \cellcolor[rgb]{1,1,0.6}{!err} & 2.65 & 4.22 & 5.29 & 5.95 & \cellcolor[rgb]{1,1,0.6}{!err} & \cellcolor[rgb]{1,1,0.6}{!err} & 4.36 & 6.18 & 5.32 & \cellcolor[rgb]{1,1,0.6}{!err} & 2.60 & 4.00 & 6.05 & 7.38 & \cellcolor[rgb]{1,1,0.6}{!err} & 2.45 & 4.95 & 6.67 & 7.67 \\
Calib (Acc) & 1.16 & $W_f$ & 2.51 & 4.01 & 5.18 & 6.09 & 6.67 & 1.86 & 2.66 & 5.30 & 6.88 & 6.11 & 2.74 & 3.98 & 5.01 & 6.76 & 7.97 & 2.59 & 3.88 & 5.79 & 7.32 & 8.24 \\
Fitts & 1.33 & $W_f$ & \cellcolor[rgb]{1,1,0.6}{!err} & 3.02 & 4.46 & 5.49 & 6.12 & \cellcolor[rgb]{1,1,0.6}{!err} & 0.386 & 4.60 & 6.35 & 5.51 & 0.758 & 2.98 & 4.26 & 6.22 & 7.52 & \cellcolor[rgb]{1,1,0.6}{!err} & 2.85 & 5.16 & 6.83 & 7.80 \\
Random $A$ & 1.27 & $W_f$ & 1.35 & 3.41 & 4.73 & 5.71 & 6.32 & \cellcolor[rgb]{1,1,0.6}{!err} & 1.61 & 4.86 & 6.54 & 5.73 & 1.74 & 3.37 & 4.54 & 6.42 & 7.69 & 1.49 & 3.25 & 5.39 & 7.01 & 7.96 \\

\end{tabular}
}%end small
\bigskip\centering
\end{table}
\renewcommand{\arraystretch}{1.0}%height of table row

\renewcommand{\arraystretch}{1.5}%height of table row
\begin{table}[t]
\caption{Model fitness results for the 2D tasks. The yellow cells indicate the best fit for each criterion. The light-blue cells indicate the candidate best-fit models (i.e., those whose $\mathit{AIC}$ and $\mathit{BIC}$ differences from the best model are less than 10).}
\label{tab:2D_fit}
\scalebox{0.85}{%use small table
\begin{tabular}{l | l || c|c|c|c|c|| c|c|c}
\hline

\multicolumn{1}{c|}{Description} & \multicolumn{1}{c||}{$\mathit{ID}$ formulation} & $R^2$ & adj. $R^2$ & $\mathit{AIC}$ & $\mathit{BIC}$ & $\mathit{RMSE}$ & $a$ & $b$ & $c$ \\ \wcline{1-10}
\#1 Baseline & $\log_{2}\left(A/W+1\right)$ & 0.9904 & \cellcolor[rgb]{1,1,0.6}{0.9899} & \cellcolor[rgb]{1,1,0.6}{147.0} & \cellcolor[rgb]{1,1,0.6}{149.0} & \cellcolor[rgb]{1,1,0.6}{10.17} & 109.7 & 99.57 & --- \\ \hline
\#2 $\mathit{ID}_e$ & $\log_{2}\left(A/W_e+1\right)$ & 0.7317 & 0.7168 & 213.7 & 215.7 & 50.79 & 22.86 & 147.2 & --- \\ \hline
\#3 Param. Opt. ($W_e$, no sqrt) & $\log_2 \left( \frac{A}{ W_e - c} +1 \right)$ & 0.9400 & 0.9330 & 185.7 & 188.7 & 25.77 & -1.399 & 108.0 & 4.026 \\ \hline
\#4 Param. Opt. ($W_e$, sqrt) & $\log_2 \left( \frac{A}{ \sqrt{ W_e^2 - c^2}} +1 \right)$ & 0.9341 & 0.9263 & 187.6 & 190.6 & 30.31 & 35.29 & 119.1 & 4.825 \\ \hline
\#5 Param. Opt. ($W$, no sqrt) & $\log_2 \left( \frac{A}{ W - c} +1 \right)$ & \cellcolor[rgb]{1,1,0.6}{0.9905} & 0.9893 & \cellcolor[rgb]{0.7,1,1}{149.0} & \cellcolor[rgb]{0.7,1,1}{151.9} & 42.29 & 110.5 & 99.08 & 0.02535 \\ \hline
\#6 Param. Opt. ($W$, sqrt) & $\log_2 \left( \frac{A}{ \sqrt{ W^2 - c^2}} +1 \right)$ & \cellcolor[rgb]{1,1,0.6}{0.9905} & 0.9893 & \cellcolor[rgb]{0.7,1,1}{148.9} & \cellcolor[rgb]{0.7,1,1}{151.9} & 11.05 & 110.7 & 99.12 & 0.2850 \\ \hline
\#7 Calib. (Acc) (given $\sigma_\mathrm{a}$) & $\log_2 \left( \frac{A}{ \sqrt{2 \pi e (\sigma_\mathrm{obs}^2 - \sigma_\mathrm{a}^2)} } +1 \right)$ & 0.9340 & 0.9303 & 185.6 & 187.6 & 25.70 & 33.14 & 120.1 & --- \\
 
\end{tabular}
}%end small
\bigskip\centering
\end{table}
\renewcommand{\arraystretch}{1.0}%height of table row

\section{Reanalyses of Previous Studies}
\label{sec:reanalysis}
Here, we reanalyze three sets of data reported in previous studies: Woodward et al.'s study using circular targets \cite{Woodward20avi} and Bi et al.'s 1D and 2D targets \cite{Bi13a}.
They conducted finger calibration tasks to obtain $\sigma_\mathrm{a}$.
The results are summarized in Table~\ref{tab:Related_fit}.
We also examined using $\sigma_\mathrm{a}$ obtained by the intercept method, but the mathematical error occurred for all three data sets.
Thus, we report the fits for Models \#1 to \#7 used in our 2D data analysis.
Note that Model \#7 is an exception: our $\sigma_\mathrm{a}$ was obtained from the finger calibration task with the ``concentrate on accuracy'' instruction, while Bi et al. used the ``rapid and accurate'' instruction and Woodward et al.'s instruction was unclear from their paper.

\renewcommand{\arraystretch}{1.5}%height of table row
\begin{table}[t]
\caption{Model fitness results for the previous studies. The yellow cells indicate the best fit for each criterion in each study. The light-blue cells indicate the candidate best-fit models (i.e., those whose $\mathit{AIC}$ and $\mathit{BIC}$ differences from the best model are less than 10). Probably because of round-off errors reported in those papers, the $R^2$ values are not exactly the same as in our analyses.}
\label{tab:Related_fit}
\scalebox{0.635}{%use small table
\begin{tabular}{l | l || c|c|c|c|c|| c|c|c|c|c|| c|c|c|c|c}
\hline

\ & & \multicolumn{5}{c||}{Woodward et al. \cite{Woodward20avi}} & \multicolumn{5}{c||}{Bi et al. \cite{Bi13a}, 1D} & \multicolumn{5}{c}{Bi et al. \cite{Bi13a}, 2D} \\
\multicolumn{1}{c|}{Description} & \multicolumn{1}{c||}{$\mathit{ID}$ formulation} & $R^2$ & adj. $R^2$ & $\mathit{AIC}$ & $\mathit{BIC}$ & $\mathit{RMSE}$ & $R^2$ & adj. $R^2$ & $\mathit{AIC}$ & $\mathit{BIC}$ & $\mathit{RMSE}$ & $R^2$ & adj. $R^2$ & $\mathit{AIC}$ & $\mathit{BIC}$ & $\mathit{RMSE}$ \\ \wcline{1-17}
\#1 Baseline & $\log_{2}\left(A/W+1\right)$ & 0.928 & 0.924 & \cellcolor[rgb]{0.7,1,1}{52.7} & \cellcolor[rgb]{0.7,1,1}{52.3} & 20.2 & 0.956 & 0.953 & \cellcolor[rgb]{0.7,1,1}{46.2} & \cellcolor[rgb]{0.7,1,1}{45.8} & \cellcolor[rgb]{1,1,0.6}{11.8} & 0.849 & 0.840 & 52.2 & 51.8 & 18.0 \\ \hline
\#2 $\mathit{ID}_e$ & $\log_{2}\left(A/W_e+1\right)$ & 0.0766 & 0.0253 & 68.0 & 67.6 & 66.3 & 0.859 & 0.851 & \cellcolor[rgb]{0.7,1,1}{53.2} & \cellcolor[rgb]{0.7,1,1}{52.7} & 18.2 & 0.789 & 0.777 & 54.2 & 53.8 & 20.4 \\ \hline
\#3 Param. Opt. ($W_e$, no sqrt) & $\log_2 \left( \frac{A}{ W_e - c} +1 \right)$ & 0.215 & 0.122 & 69.0 & 68.4 & 77.0 & \cellcolor[rgb]{1,1,0.6}{0.967} & \cellcolor[rgb]{1,1,0.6}{0.963} & \cellcolor[rgb]{0.7,1,1}{46.5} & \cellcolor[rgb]{0.7,1,1}{45.9} & 40.5 & \cellcolor[rgb]{1,1,0.6}{0.981} & \cellcolor[rgb]{1,1,0.6}{0.979} & \cellcolor[rgb]{1,1,0.6}{41.7} & \cellcolor[rgb]{1,1,0.6}{41.1} & 36.3 \\ \hline
\#4 Param. Opt. ($W_e$, sqrt) & $\log_2 \left( \frac{A}{ \sqrt{ W_e^2 - c^2}} +1 \right)$ & 0.214 & 0.122 & 69.0 & 68.4 & 71.5 & 0.960 & 0.955 & \cellcolor[rgb]{0.7,1,1}{47.6} & \cellcolor[rgb]{0.7,1,1}{47.0} & 21.8 & 0.978 & 0.975 & \cellcolor[rgb]{0.7,1,1}{42.7} & \cellcolor[rgb]{0.7,1,1}{42.1} & 15.0 \\ \hline
\#5 Param. Opt. ($W$, no sqrt) & $\log_2 \left( \frac{A}{ W - c} +1 \right)$ & \cellcolor[rgb]{1,1,0.6}{0.975} & \cellcolor[rgb]{1,1,0.6}{0.972} & \cellcolor[rgb]{1,1,0.6}{48.5} & \cellcolor[rgb]{1,1,0.6}{47.8} & 16.2 & 0.956 & 0.950 & \cellcolor[rgb]{0.7,1,1}{48.2} & \cellcolor[rgb]{0.7,1,1}{47.6} & \cellcolor[rgb]{1,1,0.6}{11.8} & 0.849 & 0.831 & 54.2 & 53.6 & 18.0 \\ \hline
\#6 Param. Opt. ($W$, sqrt) & $\log_2 \left( \frac{A}{ \sqrt{ W^2 - c^2}} +1 \right)$ & 0.971 & 0.967 & \cellcolor[rgb]{0.7,1,1}{49.3} & \cellcolor[rgb]{0.7,1,1}{48.7} & \cellcolor[rgb]{1,1,0.6}{15.6} & 0.956 & 0.950 & \cellcolor[rgb]{0.7,1,1}{48.2} & \cellcolor[rgb]{0.7,1,1}{47.6} & \cellcolor[rgb]{1,1,0.6}{11.8} & 0.849 & 0.831 & 54.2 & 53.6 & 18.0 \\ \hline
\#7 Calib (R\&A) (given $\sigma_\mathrm{a}$) & $\log_2 \left( \frac{A}{ \sqrt{2 \pi e (\sigma_\mathrm{obs}^2 - \sigma_\mathrm{a}^2)} } +1 \right)$ & 0.213 & 0.169 & 67.1 & 66.6 & 58.8 & 0.958 & 0.955 & \cellcolor[rgb]{1,1,0.6}{45.9} & \cellcolor[rgb]{1,1,0.6}{45.5} & 13.7 & 0.968 & 0.966 & \cellcolor[rgb]{0.7,1,1}{42.9} & \cellcolor[rgb]{0.7,1,1}{42.5} & \cellcolor[rgb]{1,1,0.6}{7.78} \\

\end{tabular}
}%end small
\bigskip\centering
\end{table}
\renewcommand{\arraystretch}{1.0}%height of table row

For Woodward et al.'s data, the model fitness for the baseline (Model \#1) was higher than the $\mathit{ID_e}$ (Model \#2), and Models \#3 and \#4 using $W_e$ and $c$ partially improved the fit (adjusted $R^2$ increased from 0.0253 to 0.122); these results are consistent with ours.
When the nominal $W$ and $c$ were used (Models \#5 and \#6), the fitness was improved over the baseline (\#1), and the difference was \textit{positive} according to the $\mathit{BIC}$ metric.
Lastly, use of the given $\sigma_\mathrm{a}$ value (Model \#7) gave the best fit among the $W_e$-based candidates (Models \#2, \#3, \#4, and \#7), but the fit was worse than the models using \HL{the} nominal $W$ (\#1, \#5, and \#6).
We conclude that Models \#5 and \#6 are the best choices for this dataset.

For Bi et al.'s 1D task results, Model \#7 was the best for the $\mathit{AIC}$ and $\mathit{BIC}$, which is a unique outcome among all the analyses in this paper.
Still, the differences in \HL{the} $\mathit{AIC}$ and $\mathit{BIC}$ \HL{were} not large for all models except \#2.
For Models \#5 and \#6, the $c$ value was determined as $\sim$0 to maximize the $R^2$; thus, while the $R^2$ values were the same as for Model \#1, the adjusted $R^2$, $\mathit{AIC}$, and $\mathit{BIC}$ were worse than the baseline because of the additional free parameter.
If researchers value the prediction accuracy for unknown task conditions, the $\mathit{RMSE}$ is important, and the models using \HL{the} nominal $W$ (\#1, \#5, and \#6) \HL{are} better than \#7.

For Bi et al.'s 2D task results, among all models, the best fit was shown by Model \#3, and this was significantly better than the baseline (Model \#1).
Among all the data analyses in this paper, only this case showed that a model using $W_e$ gave a better fit than those using the nominal $W$.
However, according to \HL{the} $\mathit{RMSE}$, the prediction accuracy of \HL{Model} \#3 is the worst, and thus we cannot recommend using this model for predicting $\mathit{MT}$s for new task conditions.
Rather, \HL{Model} \#4 (using $W_e$ and $c$) or \#7 (using the preset $\sigma_\mathrm{a}$) would be better.

Through these reanalyses, we reconfirmed that using a preset $\sigma_\mathrm{a}$ \HL{is} not a robust method, because \HL{$\sigma_\mathrm{a}$ can be greater than some $\sigma_\mathrm{obs}$ values and thus cause the mathematical error.}
Meanwhile, models using $c$ show the best or sub-optimal model fitness.
\HL{These results demonstrate the benefit of introducing an additional free parameter $c$ regardless of whether the nominal $W$ or $W_e$ is used.}

\section{General Discussion}
\subsection{(In)Consistency of Results from Previous FFitts Law Studies}
The first inconsistency with the previous studies is that we sometimes could not use FFitts law with $\sigma_\mathrm{a}$ because of the mathematical error.
For the 1D task, we could not use it for any derivations (finger calibrations and the intercept method; see Table~\ref{tab:1D_W_f}).
For the 2D task, only the $\sigma_\mathrm{a}$ computed from the finger calibration task with the ``concentrate on accuracy'' instruction could be used (Table~\ref{tab:2D_W_f}).
This clearly shows a limitation of the conventional FFitts law: because it depends on both the $\sigma_\mathrm{a}$ and $\sigma_\mathrm{obs}$ values, we cannot often use this methodology.

Even when we applied FFitts law with $\sigma_\mathrm{a}$ to the 2D results, the model fitness was significantly degraded as compared with the baseline model (Models \#7 vs. \#1 in Table~\ref{tab:2D_fit}).
This is inconsistent with the finding of Bi et al. \cite{Bi13a} but consistent with that of Woodward et al. on FFitts law for children whose ages ranged from 5 to 10 years \cite{Woodward20avi}.
While Woodward et al. assumed that the reason for this lower fit was the children's motor development (e.g., not precisely following the known speed-accuracy tradeoff behavior), we observed that their finding on lower model fitness also held for adults in their twenties.

As for introducing an additional free parameter $c$, we found a benefit as reported by Ko et al.\cite{Ko20FF2D}.
For the 2D task, while Model \#2 ($\mathit{ID_e}$) showed adjusted $R^2 = 0.72$, using $c$ improved the fitness: Models \#3 and \#4 showed adjusted $R^2 = 0.93$, and the $\mathit{AIC}$ and $\mathit{BIC}$ differences were significant (Table~\ref{tab:2D_fit}).
Because comparing different user groups or devices requires using $W_e$ to normalize the speed-accuracy biases, and because the effective width method assumes that the data follows Fitts' law, this additional parameter $c$ for finger tremor helps enable a more appropriate comparison.
For the 1D condition, however, the benefit was not that clear.
According to the result of the $\mathit{BIC}$ comparison, the use of Model \#2 \HL{rather than \#3 and \#4 is positively supported}, while there were no significant $\mathit{AIC}$ differences, and the $\mathit{RMSE}$ differences were less than 1 ms (Table~\ref{tab:1D_fit}).

From a different viewpoint, regarding whether to use $W$ or $W_e$, overall, we found that using $W$ gave a better fit.
For example, in our 1D data, Model \#1 showed a significantly better fit than \#2 (baseline vs. $\mathit{ID_e}$), \#5 was better than \#3 (the \HL{no-square-root} models using $W$ and $W_e$, respectively), and \#6 was better than \#4 (the \HL{square-root} models using $W$ and $W_e$).
Similarly, for the data of our 2D task and Woodward et al.'s experiment \cite{Woodward20avi}, almost the same conclusions were obtained.
In contrast, Bi et al.'s data \cite{Bi13a} showed some inconsistent results.
For their 1D data, \HL{the use of} $W$ with the parameter optimization method (Models \#5 and \#6) showed slightly worse $\mathit{AIC}$ and $\mathit{BIC}$ values than \#3 and \#4 using $W_e$, but the differences were not significant.
Moreover, the $\mathit{RMSE}$ values for Models \#5 and \#6 were smaller than those for \#3 and \#4.
For their 2D data, Models \#5 and \#6 showed significantly worse results than \#3 and \#4 according to \HL{the} $\mathit{AIC}$ and $\mathit{BIC}$.
In summary, although it does not always hold, \HL{the nearly consistent conclusion is that $W$ gives a better fit than $W_e$.}

\subsection{Reasons behind Mathematical Error in Square Root}
In our data and reanalyses of \HL{the} previous studies' data, we found that a predefined $\sigma_\mathrm{a}$ could be used in limited cases.
There are some possible reasons as follows.
First, although Bi et al. assumed that $\sigma_\mathrm{a}$ is constant \cite{Bi13a}, this is not well-supported: we found that $\sigma_\mathrm{a}$ was significantly affected by the computation method for our 1D data.
While there were no significant differences in our 2D data, the $\sigma_\mathrm{a}$ values changed slightly for the four sub-tasks, and only the $\sigma_\mathrm{a}$ value computed from the ``concentrate on accuracy'' calibration-task data did not cause the mathematical error (Table~\ref{tab:2D_W_f}).
This result rejects \HL{the notion} that $\sigma_\mathrm{a}$ is not affected by subjective speed-accuracy biases or task conditions.

Second, for both $\sigma_\mathrm{a}$ and $\sigma_\mathrm{obs}$, it is possible that the number of repetitions was not sufficient to observe a normal distribution.
In our finger calibration tasks and target-pointing tasks with preset and random target distances, not all data passed the normality test.
Thus, it is likely that we observed $\sigma_\mathrm{a}$ and $\sigma_\mathrm{obs}$ that were larger or smaller than the theoretical values.
However, our experimental design \HL{for} the number of repetitions met the requirement of {a} typical Fitts' law task, in which 15 or 25 trials per condition are recommended \cite{Soukoreff04}.
In particular, for our random-distance pointing tasks, we used 40 repetitions for each target size.
\HL{If this was not sufficient and we had to use a larger number of repetitions such as 100, it would be a limitation for FFitts law, because it would be very time-consuming and require more effort from participants and experimenters.}

Third, the model formulation of FFitts law \HL{has some points that could be refined.}
For example, in contrast to Bi et al.'s assumption, (1) either $\sigma_\mathrm{r}$ or $\sigma_\mathrm{a}$ is not normally distributed, and (2) the observed touch point variability is not expressed as $\sigma_\mathrm{obs}^2 = \sigma_\mathrm{r}^2 + \sigma_\mathrm{a}^2$ (Equation~\ref{eqn:sigma_sum}); thus, models other than the dual Gaussian distribution model are needed.

\HL{Any of the three reasons mentioned here could cause the mathematical error.}
\HL{At minimum}, we empirically observed the first and second reasons in both 1D and 2D conditions, i.e., inconstant $\sigma_\mathrm{a}$ values and non-normal distributions for $\sigma_\mathrm{obs}$ and $\sigma_\mathrm{a}$, respectively.
Further work is needed to resolve these issues for deriving a better prediction model \HL{of} touch-pointing operational times.

\subsection{Recommendations on Model Selection}
Among all the 1D and 2D conditions, we recommend \textit{not} using $\sigma_\mathrm{a}$, because it often causes the mathematical error in FFitts law.
Use of the parameter optimization method is convenient for both researchers and participants, because it is always applicable and less time-consuming.
In addition, by avoiding the finger calibration task with a 1-pixel target, we can use and compare Fitts' and FFitts laws by conducting only Fitts' law tasks with reasonably sized targets\HL{, e.g., 3 mm or larger.}
This enables testing of the model fitness by using data measured from (e.g.) a gamified task of tapping bubbles on the screen, as Woodward et al. did \cite{Woodward20avi}.

When researchers try to compare several conditions such as user groups and devices, models using $W_e$ are required.
In this case, we recommend using models with $c$: \#3 or \#4 showed better or similar prediction accuracy \HL{as} compared with \#2 ($\mathit{ID_e}$, the original effective width method).
We found no clear differences between Models \#3 \HL{and} \#4 (without and with \HL{the} square root, respectively), and thus the simpler version is easier to use, i.e., \#3.

When researchers seek to predict the $\mathit{MT}$s for a single user group or a single device, models with nominal $W$ are sufficient rather than measuring the endpoint distributions.
The baseline Model \#1 showed the best fitness for 1D and 2D conditions in our data.
Still, \HL{the} FFitts law models using parameter optimization (\#5 and \#6) showed comparable prediction accuracy for our data.
Regarding the reanalyses of previous studies, Models \#5 and \#6 showed good prediction accuracy, comparable with the baseline, for Bi et al.'s data, and \HL{they achieved the best performance for Woodward et al.'s data.}
Thus, overall, we found benefits to using the parameter optimization method without the critical negative effects of overfitting.
Because we found that Model \#6 (using \HL{the} square root and $c$) showed smaller $\mathit{RMSE}$ values than \#5 in several cases, such as our 2D task and Woodward et al.'s data, we recommend using \#6.

\subsection{Limitations and Future Work}
Our conclusions are limited by the task conditions that we used.
It is unclear whether our findings, e.g., on the best model and on when a mathematical error occurred, would hold under other conditions, such as operating a smartphone with a thumb and using much longer target distances.
Also, we tested only direct touch, and we need further studies to test the applicability of our conclusions when using other techniques such as offset cursors \cite{Potter88offset,Vogel07shift}.
We assume that, because \HL{the use of} an offset cursor can reduce the finger-touch ambiguity, the optimized parameter $c$ will be close to zero, and the models without the parameter optimization method should thus show good model fitness.
For the model-fitting results, we sometimes did not observe a great difference in the $\mathit{AIC}$ and $\mathit{BIC}$ values.
This prevented us from concluding which model was significantly better, because the results could easily change depending on the user group and the task parameters $A$ and $W$.
Much more data is needed to understand this point, which will inform our future work.

We consistently found that using a square root did not remarkably improve the fitness.
For example, for the 1D results reported in Table~\ref{tab:1D_fit}, Model \#3 using $W_e$ showed $R^2 = 0.9133$, and its square-root version (Model \#4) showed $R^2 = 0.9141$, which corresponds to an improvement of 0.0008.
Similarly, Model \#5 using $W$ showed $R^2 = 0.9814$, and its square-root version (Model \#6) showed 0.9815.
\HL{The four other datasets} reported in Tables~\ref{tab:2D_fit} and \ref{tab:Related_fit} also showed that the fitness improvements achieved by the square-root forms were less than 0.01.
It is known that Fitts' law and its variants typically show high model fitness for pointing tasks ($R^2$ is often close to 1), and thus, the remaining space to improve the fits is inherently small.
If we had observed an extremely poor fit, we could have examined whether applying the square root would significantly improve the model fitness.
However, we currently do not have such a dataset, and further experiments are needed to investigate this point.

Another unresolved point is the timing of when to compute the model fitness.
Following previous studies on FFitts law \cite{Bi13a,Luo14,Woodward20avi}, we examined the fit for $4_A \times 5_W = 20$ conditions.
For the effective width method, however, Soukoreff and MacKenzie stated that the $\mathit{ID_e}$ values should be calculated for each task condition for each participant; the participants' data should then be averaged last in order to compute the throughput (i.e., a unified performance metric) \cite{Soukoreff04}.
By that methodology, we should have calculated Equation~\ref{eqn:Wf_Bi} ($W_f=\sqrt{2 \pi e (\sigma_\mathrm{obs}^2 - \sigma_\mathrm{a}^2)}$) for the 20 conditions for each of the 12 participants.
This would have increased the chance to observe the mathematical error, because it would have required checking for it 240 times.
This notion indirectly supports that researchers should avoid using $\sigma_\mathrm{a}$.
According to Olafsdottir et al., there are at least 20 approaches to compute the throughput, depending on the order of aggregating the data \cite{Olafsdottir12test}.
We did not get deeply involved in this point and simply followed the previous FFitts law studies, yet it will be worth revisiting in the future.

\section{Conclusion}
We have revisited FFitts law and the inconsistencies in its methodology.
The parameter optimization method showed some advantages compared to measuring the finger tremor factor $\sigma_\mathrm{a}$, which often causes a negative value inside a square root in both our data and the data in previous studies.
Although the parameter optimization method is not always optimal in terms of the model fitness, it can always be used and can yield a better prediction accuracy than the baseline model without causing the overfitting problem.
Thus, as a takeaway message, we recommend using the parameter optimization method for touch-pointing tasks with small targets.
Still, the best-fit model could change depending on user groups and conditions, as we showed in the reanalyses of the data in previous studies.
To better understand touch-pointing performance and derive better models, we hope that researchers will report more data from touch-pointing experiments, even if the data shows that a novel model exhibits a lower fitness than the baseline or the data cannot be fitted because of mathematical errors.

We found no single conclusion on the best-fit model that can achieve the highest fitness for any dataset. \HL{Nevertheless, this is the first empirical demonstration of such a finding, through our analyses of two new datasets and three existing ones.}
In previous studies, researchers concluded that there was a single best model: Bi et al. stated that their proposed FFitts law with the finger calibration task was the best \cite{Bi13a}, Woodward et al. stated that the baseline Fitts' law was the best \cite{Woodward20avi}, and Ko et al. stated that their parameter optimization method was the best \cite{Ko20FF2D}.
However, we have shown that such conclusions do not always hold.
This is an important step toward better understanding of finger-touch user performance on mobile devices.
We believe that revisiting previously proposed methods and indicating that there is no consistent conclusion on the best-fit model are also important notions for establishing a standardized methodology and models for mobile device operation performance in the future.

\section*{ACKNOWLEDGMENTS}
We thank the reviewers of MobileHCI 2021 and PACM IMWUT for their valuable feedback.

\bibliographystyle{ACM-Reference-Format}
\bibliography{sample-base}

\end{document}